# THE STANDARD COSMOLOGY*

JOSHUA A. FRIEMAN

*NASA/Fermilab Astrophysics Center*
*Fermi National Accelerator Laboratory*
*Batavia, IL 60510 USA*

*Department of Astronomy and Astrophysics*
*University of Chicago*
*Chicago, IL 60637*

ABSTRACT

These lectures provide an introductory review of big bang cosmology. I discuss the expanding Friedmann-Robertson-Walker universe, summarizing the observational evidence which has led to its adoption as the 'standard' cosmological model and reviewing its basic properties. Subsequent lectures provide an overview of the early universe. The final lectures give an introduction to the inflationary universe, beginning with the motivating puzzles of the standard cosmology (the horizon and flatness problems) and ending with the inflationary production of quantum field fluctuations and their possible role in seeding the large-scale structure of the Universe.

## 1. Introduction

In the late 1970's and early 1980's, theoretical cosmology underwent a renaissance: extrapolating concepts from particle physics, in particular the standard electroweak gauge theory, to very high energies, a framework emerged in which one could meaningfully speculate about the evolution of the very early universe. This marriage of particle physics and cosmology led to a number of remarkable developments, including models for the generation of the baryon asymmetry, the inflationary scenario, the notion that topological defects could be created in cosmological phase transitions, and predictions for non-baryonic particle dark matter, to name just a few.

In recent years, observational cosmology has been undergoing its own rebirth. There has been an explosion of information on the large-scale clustering of galaxies from redshift, peculiar velocity, and photometric surveys gathered by ground-based telescopes (see the lectures by Kirshner in this volume). Studies of rich clusters of galaxies via their gravitational lensing effects as well as X-ray emission from hot intracluster gas have started to provide new clues to the distribution of dark matter. In addition, the recent detection of large-angle anisotropies in the cosmic microwave background radiation by the COBE satellite provides the first probe of structure on very large scales, while a series of anisotropy experiments on smaller scales now have tantalizing results. On the scale of the universe itself, there has been steady progress in attempts to measure the cosmological parameters (in particular, the age, expansion rate, and mean density) as well as the light element abundances more precisely (see the lectures by Schramm and Walker).

As a consequence of these observational advances, cosmology is becoming data-driven in an unprecedented way: theorists no longer have the luxury of untethered speculation, but must now confront their models with an impressive array of observations. There is still debate about the reliability and interpretation of much of the data, but we are definitely entering the 'scientific' age of cosmology: those theories which are sufficiently worked out are becoming increasingly falsifiable and many will stand or fall in the coming years. This is a very healthy development for the field. It is safe to say that at present the big bang framework for the large-scale evolution of the universe remains healthy, but that we still lack a standard, tested model for the origin and evolution of structure within this framework (see the lectures by Scherrer on structure formation).

In these introductory lectures, I cover only a very small portion of the many topics of recent interest in cosmology. The first chapters provide a general overview of the standard cosmology, the hot big bang model, focusing on its kinematics and dynamics, the observational evidence in its favor, and the current status of measurements of the global cosmological parameters. Subsequent chapters review the early universe, exploring aspects of the cosmic microwave background radiation and particle relics. The final section presents a brief introduction to the inflationary scenario for the very early universe and some of its observational implications. The inflationary scenario has not yet received the empirical backing to justify its inclusion in the 'standard' cosmological model; it is discussed here because it nevertheless provides a compelling theoretical framework for the very early universe, one which should be tested by observations in coming years.

Readers wishing to delve more fully into these topics would do well to move on to two textbooks which cover them in substantial detail: *Principles of Physical Cosmology*, by P. J. E. Peebles[1], and *The Early Universe*, by E. W. Kolb and M. S. Turner[2].

Before plunging in, I note that the appendix contains a brief discursion on notation and units.

## 2. The Standard Cosmology

### 2.1. Homogeneity, Isotropy, and the Cosmological Principle

The standard hot Big Bang model, based on the homogeneous and isotropic Friedmann-Robertson-Walker (FRW) spacetimes, is a remarkably successful operating hypothesis describing the evolution of the Universe on the largest scales. It provides a framework for such observations as the Hubble law of recession of galaxies, interpreted in terms of the expansion of the universe; the abundances of the light elements, in excellent agreement with the predictions of primordial nucleosynthesis; and the thermal spectrum and angular isotropy of the cosmic microwave background radiation (CMBR), as expected from a hot, dense early phase of expansion.

While homogeneity and isotropy are, strictly speaking, assumptions of the model, they rest on a strong and growing foundation of observational support. The evidence for angular isotropy on large scales comes from the smallness of the CMBR large-angle anisotropy detected by COBE [3] and FIRS [4] (the quadrupole anisotropy detected in the first year of COBE data is $(\Delta T/T)_{l=2} \simeq 5 \times 10^{-6}$), from the isotropy of radiation backgrounds at other wavelengths, as well as from the isotropy of deep galaxy and radio source counts. Note that these different sources carry information about quite disparate scales and types of matter in the Universe. For example, the APM[5] and EDSGC[6] surveys measured the angular positions of more than one million galaxies over an area covering about 10 % of the southern sky out to an effective depth of roughly 600 $h^{-1}$ Mpc. The galaxy surveys give us information about the local distribution of *luminous* matter in the universe. On the other hand, through the Sachs-Wolfe effect, the large-angle CMBR measurements directly probe the gravitational potential over length scales of order 6000 $h^{-1}$ Mpc, and are thus sensitive to the mass distribution itself – the distinction is important to keep in mind, since the evidence for *dark matter* should lead us to be wary about identifying the distribution of light with that of mass. We will return to the CMBR below and first focus on the galaxy distribution.

In a survey of galaxy angular positions to some limiting apparent brightness, the joint probability of finding two galaxies in elements of solid angle $d\Omega_1$ and $d\Omega_2$ separated by an angle $\theta$ is given by

$$dP = N^2 d\Omega_1 d\Omega_2 [1 + w_{gg}(\theta)] , \qquad (2.1)$$

where $N$ is the mean surface density of galaxies in the survey and $w_{gg}(\theta)$, the galaxy two-point angular correlation function, measures the excess probability over random of finding a galaxy pair with this separation. If galaxies are distributed isotropically on large scales, we should find $w_{gg}(\theta) \to 0$ at large angles, and the correlation function should scale in a definite way with survey depth, both of which are observed. The APM data for $w_{gg}$ are shown in Fig. 1 (from[5]). The correlation function is a power law, $w_{gg}(\theta) \sim \theta^{-0.67}$ for $\theta \lesssim 1^o$, but breaks below this power law at $\theta \sim 3^o$; for $\theta \gtrsim 6^o$, $|w_{gg}(\theta)| \lesssim 5 \times 10^{-4}$ and becomes lost in the noise. This behavior implies that the variance in the number of galaxies in a patch of fixed angular size becomes small for large patch size: in other words, averaged over sufficiently large angular scales, we see roughly the same number of galaxies (brighter than a fixed limit) per steradian in different parts of the sky.

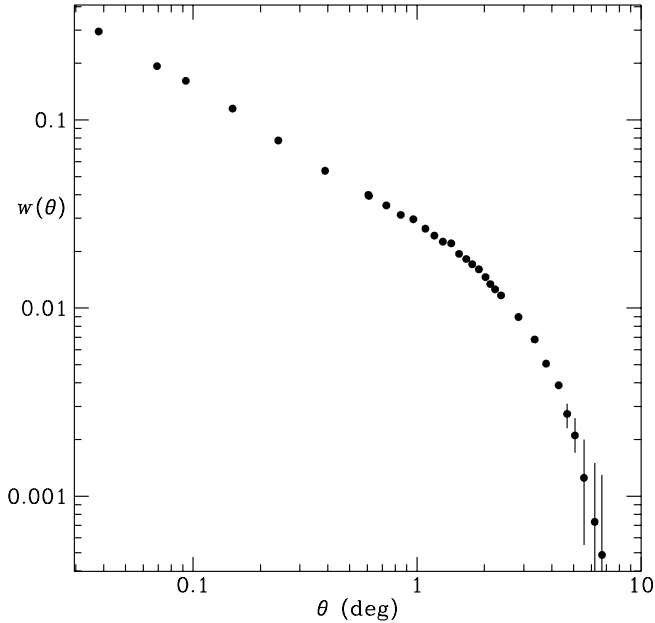

Figure 1: Galaxy angular correlation function $w_{gg}(\theta)$ measured in the APM survey, for galaxies in the apparent magnitude interval $17 < b_J < 20$.

Evidence for large-scale homogeneity comes in part from galaxy redshift surveys. However, compared to the angular photometric surveys, which currently give two-dimensional information for several million galaxies, currently complete redshift surveys yield three-dimensional information for typically several thousand galaxies in a more local neighborhood. In the redshift surveys, one can verify directly that the rms fluctuations in the spatial number density of galaxies become small when averaged over large enough scales. For example, in the full-sky surveys selected from infrared sources in the IRAS catalog (the 1.2 Jy survey of Fisher, etal.[7] and the 1-in-6 QDOT survey of Efstathiou, etal.[8] shown in Fig. 2), the rms fluctuation in the number of galaxies in cubical volumes of side $L = 60h^{-1}$ Mpc is of order $\delta N_{gal}/N_{gal} \simeq 0.2$ and decreases with increasing cell volume. This approach to homogeneity is roughly consistent with that seen in the much deeper (but two-dimensional) angular surveys. [In fact, the approach to homogeneity as a function of increasing scale observed in Fig. 1 is somewhat more gradual than was expected in the popular cold dark matter model for galaxy formation–this is the famous problem of extra large-scale power, which we will comment upon later.] Larger structures such as superclusters, great attractors, great voids, and long filaments do exist, and have received considerable attention. Of particular note in this regard was the discovery of the Great Wall, extending roughly $170 \times 60 \times 5 h^{-3}$ Mpc$^3$, by Geller and Huchra[9] in the Center for Astrophysics (CfA) survey extension. But in a statistical sense, the net fluctuations in galaxy number become small on the largest scales where they have been reliably counted in large-area redshift surveys. This is consistent with the visual impression from Fig. 2.

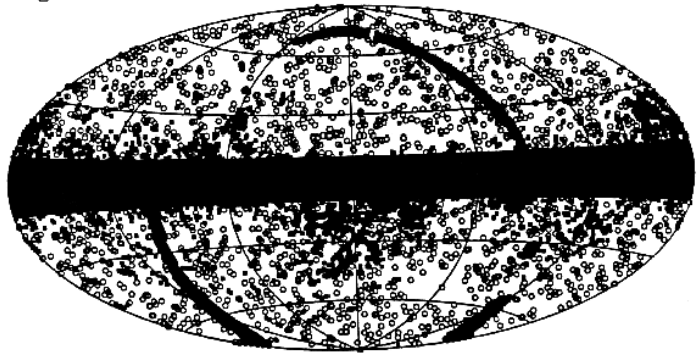

Figure 2: An equal-area projection of the IRAS-QDOT galaxies (open circles) at distances in the range $20 - 500h^{-1}$ Mpc. Solid squares are regions not surveyed, and the black band is an excluded region encompassing the Galactic disk (from ref.10).

This trend is confirmed by the visual appearance of large-scale structure in the ongoing Las Campanas redshift survey[11]: going considerably deeper than the CfA survey, they do not find evidence for coherent structures larger than the Great Wall. The Las Campanas survey uses a multifiber spectrograph to simultaneously measure many redshifts in the same field, an ingenious technological development which makes possible the extension of complete large-area redshift surveys to greater depths in a finite survey time. Advances in multi-fiber spectroscopy will be further exploited by the Sloan Digital Sky Survey, which aims to measure one million galaxy redshifts over a contiguous area of $\pi$ sr in the northern sky. This survey will use a 600-fiber spectrograph to accumulate redshifts at an unprecedented rate. A concurrent photometric survey will measure angular positions for roughly 50 million galaxies.

From the CMBR and galaxy observations discussed above, we infer that on large scales the universe appears to be distributed in a statistically homogeneous and isotropic way around us. To step from this evidence to a cosmological model, we must add to it a further assumption, because we observe the universe from only a single vantage point and have direct information only about conditions on our past light cone. This assumption, often called the *Copernican Principle*, states that we do not occupy a special position in the Universe: the conditions we observe are typical of those seen by observers on distant galaxies. By itself, the Copernican Principle would allow a large range of permissible cosmological models, e.g., an inhomogeneous fractal universe with self-similar structure on all scales, or homogeneous models with anisotropic expansion in different directions. However, when combined with the observations of isotropy above, the Copernican principle becomes quite powerful, for it implies that the Universe should appear isotropic about every point. By a straightforward geometric argument, isotropy about every point (or more precisely, about every local fundamental observer comoving with the cosmic fluid) in turn implies that the Universe is spatially homogeneous, i.e., that there exist spacelike 3-surfaces of uniform energy-momentum which evolve according to a universal time $t$. This more general assumption of global homogeneity and isotropy is called the **Cosmological Principle**, a symmetry principle that is the foundation of the standard cosmology.

At this point, one is naturally tempted to ask why the universe should have such strong symmetries of isotropy and homogeneity. The attempt to provide a dynamical answer to this question is one of the motivating forces behind the inflationary scenario, which we shall discuss later on. For now, we will adopt the cosmological principle as a working hypothesis and explore some of its consequences.

### 2.2. Hubble's Law and Particle Kinematics

The cosmological principle leads directly to Hubble's law of the expanding universe. Consider a triangle formed by three fundamental observers (each of whom defines the local cosmic rest frame) at some initial time. If the universe remains homogeneous and isotropic over time, then the triangle at all later times must be similar to the original one: the length of each side scales up by the same factor $a(t)$. Extending this argument to all other fundamental observers, we see that $a(t)$ must be a universal scale factor, such that the distance $\ell(t)$ between any two fundamental observers satisfies $\ell(t) = \ell_0 a(t)$, where $\ell_0$ is the initial separation. Then the relative speed of one observer with respect to the other is given by

$$v(t) = d\ell/dt = \dot{a}\ell_0 = (\dot{a}/a)\ell(t) \equiv H(t)\ell(t) \ . \tag{2.2}$$

Note that $H$ is in general time-dependent, but for observers sufficiently nearby that the light travel time $\ell/c$ is small compared to the time over which $H$ changes appreciably, we can replace $H$ by its present value, $H(t_0) \equiv H_0$, where subscript 0 denotes the present epoch. Relation (2.2) then reproduces Hubble's observation in 1929 that the recession speed of a galaxy is proportional to its distance from us. The proportionality constant, $H_0 = (\dot{a}/a)_0$, is now known as the Hubble parameter.

A recent example of this is shown in Fig. 3, taken from ref.[12]: each circle indicates a cluster of galaxies, with distance estimated using the Tully-Fisher relation for spiral galaxies and recession velocity inferred from the redshift (see below); the horizontal error bars indicate the $1-\sigma$ spread in inferred distances for a number of galaxies per cluster. The slope of the straight line fit to the points is $H_0 \simeq 80$ km/sec/Mpc; other methods of estimating $H_0$ yield values in the range 40 - 100 km/sec/Mpc, and it is conventional to write this as $H_0 = 100h$ km/sec/Mpc, with $0.4 < h < 1$. (The closed circles around 40 Mpc indicate clusters thought to be

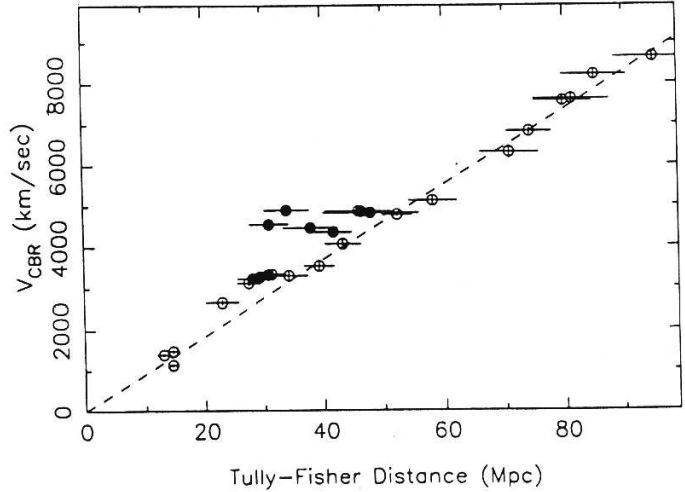

Figure 3: Hubble's law: galaxy recession speed is shown against distance inferred from the Tully-Fisher relation.

falling into the Great Attractor region, an inhomogeneity that locally perturbs the Hubble flow.)

The cosmological principle, with eqn.(2.2), tells us about the kinematics of particle motions in the universe. First consider non-relativistic particles. A massive free particle passes a fundamental observer $FO_1$ at time $t$ with relative speed $v_p \ll c$: since the FO is at rest with respect to the local fluid, $v_p$ is the peculiar velocity of the particle with respect to the local rest frame. After a time interval $dt$, the particle has moved a distance $d\ell = v_p dt$ and overtakes a second fundamental observer $FO_2$, who has a speed $dv = H(t)d\ell = Hv_p dt$ relative to $FO_1$. At that time, $FO_2$ measures the particle's peculiar velocity to be $v_p(t + dt) = v_p(t) - dv$. Thus, the peculiar velocity satisfies the equation of motion

$$\frac{dv_p}{dt} = -\frac{dv}{dt} = -Hv_p = -\left(\frac{\dot{a}}{a}\right) v_p \ . \tag{2.3}$$

The solution is

$$v_p \propto \frac{1}{a} \ , \tag{2.4}$$

that is, in a homogeneous and isotropic universe, peculiar velocities decrease with the expansion. Now consider a dilute non-relativistic gas in thermal equilibrium. By equipartition, the mean kinetic energy is $m\langle v^2 \rangle/2 = (3/2)k_B T_g$, where $T_g$ is the gas temperature; by eqn.(2.4) the gas temperature satisfies $T_g \sim v^2 \sim a^{-2}$. This applies, for example, to baryons well after they decouple from the photon background (before decoupling, the baryon temperature is tied by scattering to the photon temperature, which scales differently–see below).

The same reasoning can be applied to study the evolution of relativistic particles. Consider a photon with initial frequency $\nu(t)$ as measured by $FO_1$. After the interval $dt$, it has travelled a distance $d\ell = cdt$ and passes $FO_2$, who has a speed $v = Hd\ell = Hcdt$ with respect to $FO_1$. The photon frequency $\nu(t+dt)$ observed by $FO_2$ is given by the first-order Doppler shift,

$$d\nu = \nu(t+dt) - \nu(t) = -\nu\frac{v}{c} = -\nu H dt \ . \tag{2.5}$$

The frequency thus satisfies

$$\frac{d\nu}{dt} = -\nu\frac{\dot{a}}{a} \ , \tag{2.6}$$

with solution

$$\nu(t) \propto \frac{1}{a(t)} \ . \tag{2.7}$$

The photon frequency is redshifted with the expansion, and its wavelength is stretched with the scale factor, $\lambda(t) \sim \nu^{-1} \sim a$. For a photon gas in thermal equilibrium, the temperature thus scales inversely with the scale factor, $T_r \sim \langle E_r \rangle \sim \nu \sim a^{-1}$. Generalizing the argument, one finds that the De Broglie wavelengths of all free particles follow the scale factor in this way. This leads us to define the redshift $z$:

$$1 + z(t_e) = \frac{\lambda_{obs}}{\lambda_{em}} = \frac{a(t_0)}{a(t_e)} \tag{2.8}$$

is the ratio of the photon wavelength at emission ($t_e$) to its wavelength observed at the present ($t_0$). This expression holds more generally than this derivation suggests: it is valid even when the period $\nu^{-1}$ is comparable to or longer than $t$ (although its operational meaning is then less transparent). The redshift $z$ thus plays several roles: through the Doppler shift, it is a measure of recession velocity, $v \simeq cz$; through the Hubble law, it is a measure of distance, $v = cz = H_0\ell$; and through eqn.(2.8), the redshift $z(t)$ can be used to characterize a cosmological epoch $t$. The most distant objects directly observed are luminous quasars, which have been seen out to redshifts approaching $z = 5$. By comparison, the photons in the cosmic microwave background radiation were probably emitted (or, more precisely, last scattered) at a redshift $z \simeq 1000$.

### 2.3. The Metric: Friedmann-Robertson-Walker Models

In the context of general relativity, the cosmological principle severely restricts the form of the spatial geometry: a uniform stress-energy momentum tensor implies that the constant-time 3-surfaces have uniform spatial curvature. The metric on such surfaces has the form

$$ds_3^2 = a^2(t)\left[\frac{dr^2}{1-kr^2} + r^2\left(d\theta^2 + \sin^2\theta d\phi^2\right)\right] \ , \tag{2.9}$$

where $a(t)$ is the global scale factor which describes the overall expansion or contraction, $r, \theta, \phi$ are the (fixed) comoving coordinates carried by the fundamental observers, and $k = 0, 1, -1$ is the sign of the spatial curvature. The case $k = 0$ corresponds to flat, Euclidean 3-space ($R^3$) written in spherical coordinates, $k = 1$ corresponds to the geometry of the three-sphere ($S^3$), and $k = -1$ to the 3-hyperboloid ($H^3$), the three-dimensional analogue of a hyperbolic saddle. Thus, models with $k \leq 0$ are spatially infinite (open), while those with $k = 1$ are spatially finite (closed). (Note that the apparent singularity at $r = \pm 1$ for $k = 1$ is the usual coordinate singularity at the poles of a sphere.) By homogeneity and isotropy, the full spacetime metric then takes the Friedmann-Robertson-Walker (FRW) form

$$ds^2 = g_{\mu\nu}^{FRW} dx^\mu dx^\nu = dt^2 - ds_3^2 \quad , \tag{2.10}$$

where $t$ is proper time measured on the clocks carried by the fundamental comoving observers.

We can now be somewhat more precise about the assumption of homogeneity and isotropy: clearly this is meant to be taken in some average sense over large scales. It is useful to think of this perturbatively: the actual spacetime metric can be written as the FRW metric plus a perturbation: $g_{\mu\nu} = g_{\mu\nu}^{FRW} + h_{\mu\nu}(\mathbf{x}, t)$. In an appropriate gauge, $h_{00}$ satisfies an equation of motion analogous to the Newtonian potential $\phi$. Through the Sachs-Wolfe effect, the COBE observations roughly indicate that $\delta T/T \sim \phi \sim h_{00} \sim 10^{-5} \ll g_{00}^{FRW} = 1$: the departure from the FRW metric on large scales is very small (at least in the chosen gauge). This argument extends to smaller scales as well: although galaxies represent highly non-linear condensations of the mass density, their associated gravitational potential is small, $\phi \lesssim 10^{-4}$. Thus at least when averaged over scales larger than individual galaxies, the perturbation to the FRW spacetime metric associated with inhomogeneities is very small, and the homogeneity assumption is an excellent first approximation.

The FRW models are characterized by the global scale factor $a(t)$, whose dynamics is determined by the matter content of the universe through Einstein's equations,

$$R_{\mu\nu} - \frac{1}{2} g_{\mu\nu} R - \Lambda g_{\mu\nu} = 8\pi G T_{\mu\nu} \quad . \tag{2.11}$$

Treating the matter as a perfect fluid locally at rest, and using the FRW metric, these become the Friedmann equations

$$H^2 \equiv \left(\frac{\dot{a}}{a}\right)^2 = \frac{8\pi G}{3} \rho - \frac{k}{a^2} + \frac{\Lambda}{3} \tag{2.12}$$

and

$$\frac{\ddot{a}}{a} = -\frac{4\pi G(\rho + 3p)}{3} + \frac{\Lambda}{3} \quad . \tag{2.13}$$

Here $\rho = T_{00}$ is the mean energy density of matter, $p = T_{ii}$ is its pressure, and $\Lambda$ is the cosmological constant, *i.e.*, the effective contribution to the energy-momentum from the vacuum state. Although it follows from the Einstein equations, it is also useful to separately write down the energy-momentum conservation equation, $\nabla^\nu T_{\mu\nu} = 0$, for the fluid in the FRW universe,

$$\frac{d\rho}{dt} + 3H(p + \rho) = 0 \ . \tag{2.14}$$

Observations suggest that the fluid energy density of the universe is currently dominated by non-relativistic matter ($m$), while the early universe was dominated by ultrarelativistic particles, or radiation ($r$). Non-relativistic matter is also sometimes called pressureless dust, because its fluid pressure is dynamically negligible,

$$p_m \sim \rho_m \langle v^2 \rangle \ll \rho_m \ . \tag{2.15}$$

In this case, from eqn.(2.14), the energy density scales as

$$\rho_m \sim a^{-3} \ . \tag{2.16}$$

This is consistent with conservation of particle number in a fixed comoving volume ($n_m \sim a^{-3}$), since the particle energy is dominated by its rest-mass, $\rho_m = m n_m$. For radiation, the equation of state is

$$p_r = \rho_r/3 \ , \tag{2.17}$$

and eqn.(2.14) then yields

$$\rho_r \sim a^{-4} \ . \tag{2.18}$$

The faster fall-off of $\rho_r$ with the scale factor is consistent with the Doppler redshift of the frequency of a photon emitted by a receding observer in the expanding universe, eqn.(2.7). For massless particles with average energy $\langle E_r \rangle$, this implies $\langle E_r \rangle \sim a^{-1}$ and thus $\rho_r \simeq \langle E_r \rangle n_r \sim a^{-4}$.

Two features of the solutions to the Friedmann equation (2.12) with $\Lambda = 0$ are worth noting. In this case, there is a one-to-one correspondence between the spatial geometry and the fate of the universe: open models ($k \leq 0$) expand forever, while closed models ($k > 0$) eventually recollapse, because the energy density for matter and radiation fall off faster than $a^{-2}$. Conversely, in the early universe, $a \ll a_0$, the matter and radiation terms dominate over the spatial curvature, and the dynamics of the model is well approximated by setting $k = 0$. In this limit, for a matter-dominated universe, eqns.(2.12) and (2.16) give the solution

$$a(t) \sim t^{2/3} \quad ; \quad \rho_m = \frac{1}{6\pi G t^2} \quad (\text{MD}, k = 0) \tag{2.19}$$

while for radiation-domination, from (2.12) and (2.18)

$$a(t) \sim t^{1/2} \quad ; \quad \rho_r = \frac{3}{32\pi G t^2} \quad (\text{RD}, k = 0) \ . \tag{2.20}$$

This clearly generalizes for a fluid with equation of state $p_f = w\rho_f$:

$$\rho_f \sim a^{-3(1+w)} \rightarrow a(t) \sim t^{2/3(1+w)} \quad ; \quad \rho_f = \frac{1}{6\pi G t^2 (1+w)^2} \quad . \tag{2.21}$$

In the next few sections we focus on the cosmologically recent matter-dominated era, and subsequently turn to the radiation-dominated early universe.

## 3. Cosmological Parameters: $H_0$ and $t_0$

The principal observable cosmological parameters of the FRW models are the Hubble parameter, $H_0 = (\dot{a}/a)_0$, the age of the Universe, $t_0$, the present (non-relativistic) mass density relative to the 'critical' density of the spatially flat, Einstein-de Sitter ($k = \Lambda = 0$) model,

$$\Omega_0 = \rho_0/\rho_{crit} = 8\pi G \rho_0 / 3 H_0^2 \quad , \tag{3.1}$$

the deceleration parameter, $q_0 = -(\ddot{a}a/\dot{a}^2)_0$, which measures the rate at which the gravitational attraction of the matter is slowing down the expansion, and the contribution of the cosmological constant to the present expansion rate, $\lambda_0 = \Lambda/3H_0^2$. From the Friedmann equations, these parameters are related by

$$1 = \Omega_0 + \lambda_0 - \frac{k}{a_0^2 H_0^2} \tag{3.2}$$

$$q_0 = \frac{\Omega_0}{2} - \lambda_0 \quad . \tag{3.3}$$

For vanishing cosmological constant, $\Omega_0 - 1$ determines the sign of the spatial curvature: $\Omega_0 = 1$ for the spatially flat model ($k = 0$), and it is less than one for open models.

The age of the universe is related to the other parameters through an expression of the form $H_0 t_0 = 1.02 h (t_0/10^{10}\text{yr}) = f(\Omega_0, \lambda_0)$, where $f$ is a function of order unity. Thus the *Hubble time* $H_0^{-1} = 9.8 \times 10^9 h^{-1}$ yr sets the timescale for the age of the universe, while the *Hubble length* $cH_0^{-1} = 3000 h^{-1}$ Mpc sets the lengthscale for the present observable universe. For a matter-dominated universe with $\Lambda = 0$, $f$ falls monotonically with increasing $\Omega_0$, and two useful limits are $f(0,0) = 1$ ($a \sim t$ for $\Omega_0 = 0$) and $f(1,0) = 2/3$ (since $a \sim t^{2/3}$ for $\Omega_0 = 1$). More generally, over the range $0 < \Omega_0 \leq 1$, $k \leq 0$, $\Omega_0 - 3\lambda_0/7 \leq 1$, an excellent approximation is[13]

$$H_0 t_0 \simeq \frac{2}{3} \frac{\sinh^{-1}(\sqrt{(1-\Omega_a)/\Omega_a})}{\sqrt{1-\Omega_a}} \tag{3.4}$$

where

$$\Omega_a = \Omega_0 - 0.3(\Omega_0 + \lambda_0) + 0.3 \quad . \tag{3.5}$$

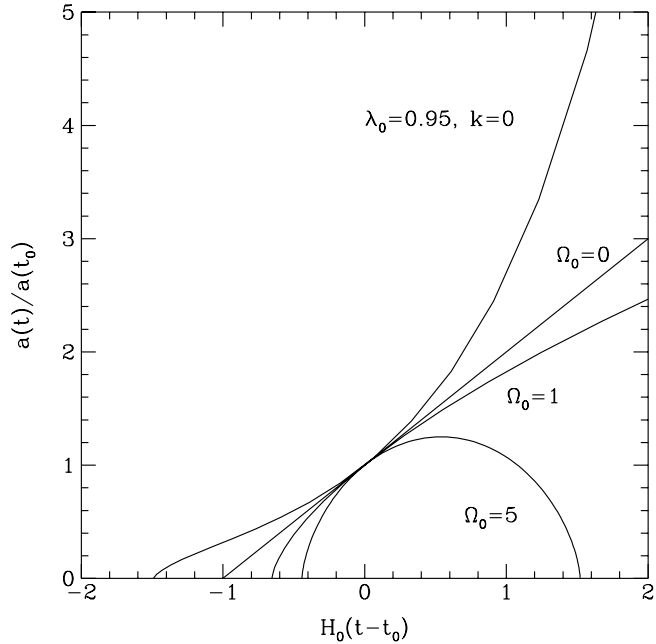

Figure 4: Evolution of the cosmic scale factor (scaled to its present value).

Some examples of the evolution of the scale factor $a(t)$ are shown in Fig.4, including three cases with $\Lambda = 0$: empty ($\Omega_0 = 0$), flat ($\Omega_0 = 1$), and closed ($\Omega_0 = 5$), and one example of a flat ($k = 0$) model dominated by a cosmological constant ($\lambda_0 = 0.95 = 1 - \Omega_0$). This figure displays the decrease in the expansion age $H_0 t_0$ from unity as $\Omega_0$ is increased from zero. It also demonstrates that a significant cosmological constant term can make $H_0 t_0$ large. The solution for the spatially flat ($k = 0$) model with non-zero $\Lambda$ has the form

$$\frac{a(t)}{a(t_0)} = \left(\frac{\Omega_0}{\lambda_0}\right)^{1/3} \sinh^{2/3}\left[\frac{3\lambda_0^{1/2} H_0 t}{2}\right] \quad , \tag{3.6}$$

where $\lambda_0 + \Omega_0 = 1$. At late times, it approaches the exponential de Sitter solution,

$$a(t) \sim e^{\sqrt{\frac{\Lambda}{3}} t} \quad . \tag{3.7}$$

Much early effort was spent trying to measure or constrain the parameters $q_0$ and $\lambda_0$ through the classical 'cosmological tests', such as the Hubble diagram, angular size as a function of redshift, and galaxy counts as a function of redshift and apparent brightness. For example, to construct the Hubble diagram, one measures the apparent brightness of a well-defined sample of objects (say, the brightest galaxies in clusters) as a function of the object's redshift; for galaxies of fixed intrinsic luminosity, the scaling of apparent magnitude with redshift is a function of the cosmological parameters. Unfortunately, galaxies at large distances, where the distinction between model parameters becomes observable, are seen when they were much younger than their nearby counterparts, so a model for galaxy luminosity evolution must be used to interpret the results. Significant progress has been made in understanding galaxy evolution, and there is hope that the effects of evolution and cosmology might be disentangled in coming years, but these tests currently do not place stringent universally accepted constraints on the cosmological parameters. Recently, it has been pointed out[14] that the probability that a quasar at a given redshift is gravitationally lensed by a foreground galaxy is a sensitive test for the cosmological constant: in models with $\lambda_0 > 0$, one generally expects a higher lensing probability. Based on surveys for lensed quasars, the bound $\lambda_0 \lesssim 0.8$ has been inferred in the case of a spatially flat ($k = 0$) universe[15]. For more on the standard cosmological tests, the reader is referred to Peebles[1].

### 3.1. $H_0$ and the Distance Scale

The Hubble parameter relates the observed recession velocity $v_r$ or redshift $z$ of a galaxy to its distance $d$: for $v_r \ll c$, the recession velocity is $v_r = cz = H_0 d + v_p$, where $v_p$ is the peculiar radial velocity of the galaxy with respect to the Hubble flow, usually assumed to arise from gravitational clustering. Galaxy redshifts can be measured quite accurately, so all the difficulty in determining $H_0$ resides in finding reliable distance indicators for extragalactic objects at distances large enough that the Hubble term dominates over the peculiar motion. Observed peculiar velocities are typically of order 300 km/sec, so that distance measurements beyond 40 Mpc or more (recession velocities above 4000 km/sec) are required for reasonable accuracy.

A wide variety of techniques has been used to establish an extragalactic distance scale[16], and this is reflected in the spread of results for $H_0$, roughly $40 - 100$ km/sec/Mpc. Distance estimates made using methods such as the Tully-Fisher relation between 21-cm rotation speed and infrared luminosity for spiral galaxies, calibrated by observations of Cepheid variable stars in several nearby galaxies, have yielded high values for the expansion rate, roughly $H_0 = 80 \pm 10$ km/sec/Mpc. Two newer methods, planetary nebula luminosity functions[17] and galaxy surface brightness fluctuations[18] yield values for $H_0$ in this range as well, and are being further developed. On the other hand, methods using Type Ia supernovae as standard candles have yielded low values, $H_0 \simeq 50 \pm 10$ km/sec/Mpc. SNe Ia are thought to be the explosions of white dwarfs which accrete matter from binary companions until they reach the Chandrasekhar mass, and there is some evidence that they form a homogeneous class; they also have the advantage that they can be observed to great distances. In the future, Hubble Space Telescope observations of Cepheids in other nearby galaxies (and perhaps as far as the Virgo cluster) which are hosts to SNe Ia or which can be used as Tully-Fisher calibrators should help improve the situation. The recent discovery of a probable Type Ia supernova at $z = 0.46$[19] also raises hopes that a sample of SNe Ia at $z \sim 0.5$ could significantly constrain $q_0$, provided the dispersion in SNe Ia luminosities is sufficiently narrow.

There are also a variety of methods being employed to measure the distances of extragalactic objects directly, bypassing the extragalactic distance ladder built up from Cepheids. Using the expanding photosphere method, Schmidt, Kirshner, and Eastman[20] have determined the distances to 10 type II supernovae at large distances, and find good agreement with the Tully-Fisher distances for these galaxies. Other 'direct' methods which hold future promise include measurement of the Sunyaev-Zel'dovich effect, due to the Compton upscattering of CMBR photons by hot gas in rich clusters[21], and the differential time delay between images in gravitationally lensed quasars[22].

### 3.2. The Age of the Universe, $t_0$

Three methods have traditionally been used to infer the age of the Universe, $t_0$. Nuclear cosmochronology is based on radioactive dating of $r-$process elements, that is, heavy elements formed by rapid neutron capture, most probably in supernovae. The element ratios Re/Os and Ur/Th generally indicate $t_0 = 10 - 20$ Gyr[23], with a large uncertainty due to the unknown element formation history (e.g., the star formation rate over time). A second method involves the cooling of white dwarfs: when low-mass stars exhaust their nuclear fuel, they become degenerate white dwarfs, gradually cooling and becoming fainter. The number of white dwarfs as a function of luminosity drops dramatically for $L_{wd} < 3 \times 10^{-5} L_\odot$, suggesting that there has not been sufficient time for them to fade below this value. Coupled with models of white dwarf cooling, this implies that the age of the galactic disk is about $t_0 \simeq 10 \pm 2$ Gyr[24].

The most extensively studied technique for constraining $t_0$ is the determination of the ages of the oldest globular clusters in the galaxy. When stars finish burning hydrogen, they turn off the main sequence, characteristically reddening and brightening. By observing the color-magnitude (color vs. apparent brightness) diagram for a cluster, one can determine the apparent brightness of stars in the cluster that are now leaving the main sequence. Knowing the distance to the cluster then gives the absolute luminosity of stars at the turn-off. On the other hand, stellar evolution theory relates stellar luminosity to the time a star spends on the main sequence: crudely, a star turns off the main sequence when a fixed fraction $f \simeq 0.15$ of its initial hydrogen mass $M_H = X_H M$ has been converted into helium (here $X_H \sim 0.75$ is the initial H mass fraction of the star). The conversion of H to He releases $6.4 \times 10^{18}$ erg for every gm of H burned. Thus, over the course of the main sequence, the total energy radiated by a star is $E \propto f X_H M \simeq 10^{52} f X_H (M/M_\odot)$ erg, and its evolutionary lifetime on the main sequence is $\tau_{ms} \simeq E/L \simeq 10^{11} f X_H (M/M_\odot)/(L/L_\odot)$ yr. Now, stellar models indicate that the relationship between mass and main sequence luminosity is roughly $L \sim M^4$, yielding a main-sequence lifetime of $\tau_{ms} \simeq 1.2 \times 10^{10} (L/L_\odot)^{-3/4}$ years. Thus, the measurement of the main sequence turn-off luminosity gives an estimate of the age of the cluster. The turn-off luminosity in the oldest globular clusters in the galaxy is slightly below that of the sun, yielding the age estimate $t_{gc} = (13 - 15) \pm 3$

Gyr[25]. This argument is a gross oversimplification of the complex process of stellar modelling and cluster isochrone fitting, but it gives a heuristic understanding of the resulting age estimate. The largest source of error is apparently the uncertainty in the distances to the globular clusters. It is hoped that observations with the corrected Hubble Space Telescope mirror, successfully repaired in late 1993, could reduce the uncertainty in $t_{gc}$ to as little 10%. There may also be residual systematic model uncertainties associated with the initial hydrogen (or helium) fraction $X_H$, opacities, metallicity, etc.

We now turn to the final cosmological parameter of interest, the density of the universe.

## 4. Cosmological Parameters: $\Omega_0$ and Dark Matter

It is convenient to parameterize the mass density of the universe in terms of the mass–to–light ratio, say in the $V_T$ band, $\Upsilon = \langle M/L \rangle / (M_\odot/L_\odot)$. Dividing the present critical density,

$$\rho_c = 3H_0^2/8\pi G = 1.88 h^2 \times 10^{-29} \text{gm cm}^{-3} = 2.8 \times 10^{11} h^2 M_\odot \text{ Mpc}^{-3} \qquad (4.1)$$

by the observed mean luminosity density $j_{V_T} \simeq 2.4 \times 10^8 h L_\odot \text{ Mpc}^{-3}$, the critical mass–to–light ratio for the $\Omega_0 = 1$ universe is $\Upsilon_c \simeq 1200h$, and the cosmic density parameter can be expressed as $\Omega_0 = 8 \times 10^{-4} \ h^{-1} \Upsilon$. The mass–to–light ratio in the solar neighborhood is approximately $\Upsilon = 5$, while the central cores of elliptical galaxies yield $\Upsilon \simeq 12h$, so the density of luminous matter (that is, of matter associated with typical stellar populations) is inferred to be $\Omega_{lum} \sim 0.007$. However, it is well known that the luminous parts of galaxies are not the whole story: there is strong evidence from flat spiral galaxy rotation curves, from a variety of observations of galaxy clusters, and from large-scale peculiar motions that there is a substantial amount of dark matter in the universe.

### 4.1. Dark matter in galaxies

An example of this phenomenon for galaxies is shown in Fig. 5, which shows a model for the rotation speed of the galaxy as a function of radius. The three lower curves show the contributions from the luminous components (two bulge components and the disk[26]). Clearly, they cannot account for the observed rotation speed beyond a few kpc from the galactic center. To match the observations (taken from the compilation in ref.[27]), one assumes in addition a quasi-spherical distribution of dark matter, called the halo, with a density profile of the form

$$\rho_h(r) \propto \frac{1}{a^2 + r^2} \quad . \qquad (4.2)$$

In Fig. 5, the adopted halo core radius is $a = 3.5$ kpc. On large scales, the halo mass scales as $M_h(r) \sim r$, yielding a flat rotation curve, $v_c^2 = GM(r)/r \to 220$ km/sec for the model shown in Fig. 5. Clearly, eqn.(4.2) cannot extend to arbitrarily large radii, because the enclosed mass would diverge; the halo is presumably truncated by tidal or other effects at large $r$. The observation of high proper motion stars in the solar neighborhood (presumably bound to the Galaxy) implies that the local value of the galactic escape velocity exceeds 450–500 km/sec, and this can be used to place a lower bound on the halo truncation radius. If the halo is sharply truncated at $r_t$, the escape speed $v_e$ at $r < r_t$ satisfies[28]

$$v_e^2 = 2v_c^2(1 + \ln(r_t/r)] \quad . \tag{4.3}$$

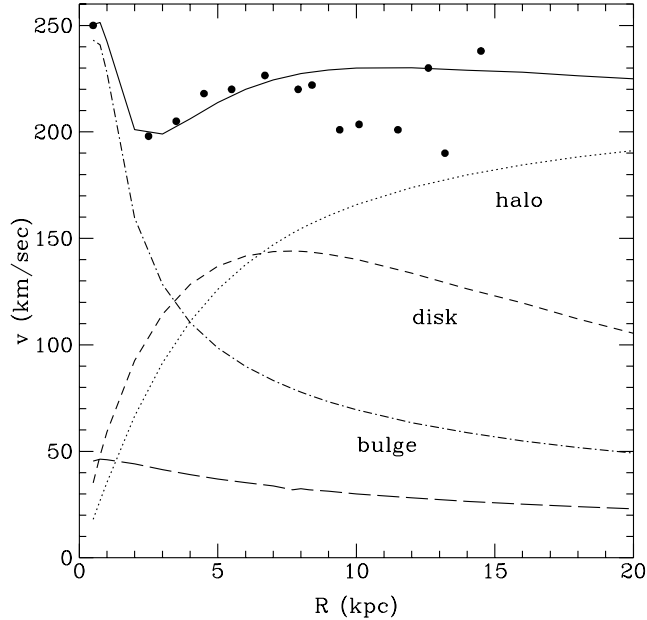

Figure 5: Rotation curve for our galaxy: an average over a number of different observations (solid points) is shown with a model that includes luminous and dark matter.

Using $v_e \geq 500$ km/sec at the solar radius $r = R_0 = 8.5$ kpc and $v_c = 220$ yields $r_t \gtrsim 40$ kpc and a total galaxy mass $M_{gal} \gtrsim 5 \times 10^{11} M_\odot$. Comparing with the luminosity of the disk plus the bulge, $L_{gal} = 1.4 \times 10^{10} L_\odot$ in the V band, this implies that the total mass–to–light ratio for the Milky Way is at least $\Upsilon_{MW} \gtrsim 35$. This is consistent with the requirement that distant globular clusters and satellite galaxies are bound to the Galaxy, as well as with mass–to–light ratios inferred from more extended flat rotation curves in other spiral galaxies[27]. If these systems are typical of the universe, we infer $\Omega_0 \gtrsim 0.02 h^{-1}$ for the matter associated with galaxy halos.

It is interesting to compare these values with the baryon density $\Omega_B$ inferred from primordial nucleosynthesis (see the lectures by Schramm and Walker), which[29][30] has been restricted to the range $0.010 < \Omega_B h^2 < 0.015$. Comparison with $\Omega_{lum}$ above suggests that some or most of the baryons are dark or in underluminous populations. Furthermore, comparison with the escape speed and rotation curve bound $\Omega_0 \gtrsim 0.02h^{-1}$ shows that baryons could constitute some or all of the dark matter in galaxy halos, depending on their extent.

One possibility for dark baryons in halos would be degenerate brown dwarfs, substellar ($M < 0.08 M_\odot$) objects which did not reach sufficiently high temperature to burn hydrogen. Currently several independent groups are searching for halo dwarfs (which have been dubbed MACHOS, for massive compact halo objects); the signature is a microlensing event, in which a background star, say in the LMC or the bulge of the Milky Way, symmetrically brightens and fades as a MACHO passes near its line of sight.[31] Although they are distinguishable from ordinary variable and flare stars by the time symmetry and achromaticity of the light curve, such microlensing events would be intrinsically rare–the probability that a given background star is lensed is $\tau \sim (v_c/c)^2 \sim 5 \times 10^{-7}$, so a large number of stars must be accurately monitored. The American-Australian MACHO project currently obtains CCD photometry for several million stars per night, and has discovered many periodic variable stars. The French EROS and Polish OGLE groups are also making progress. In September and October 1993, all three groups reported the first evidence for microlensing events by objects in the Galaxy[32,33,34]. The most probable masses for the lensing objects are roughly in the range $0.01 - 0.5 M_\odot$. As data is accumulated over the next few years, we will be able to infer information about the contribution of MACHOs to the Milky Way halo. It is also worth noting that microlensing in external galaxies has been sought previously by monitoring the brightness of gravitationally lensed QSO images over time. In particular, a microlensing event in the lensed quasar 2237+0305 has been established,[35] but in this case it is difficult to establish the MACHO mass, since the microlensing probability is closer to unity.

### 4.2. Dark matter in clusters

Moving to larger scales, for clusters of galaxies the traditional dynamical method of estimating cluster masses and mass–to–light ratios, first used by Zwicky who discovered the 'missing mass' problem in the 1930's, has been to apply the virial theorem to measured cluster velocity dispersions, $M_{tot} \propto \langle v^2 \rangle / \langle R_{ij}^2 \rangle$, where $\langle v^2 \rangle$ is the velocity dispersion of the galaxies in a cluster and $R_{ij}$ is the separation between them. This method assumes that galaxies trace the cluster mass and that the galaxy velocity distribution is isotropic, both of which may not be good approximations. Independent information on the dark matter distribution in the inner cores of clusters comes from the giant luminous arcs and arclets, high redshift galaxies gravitationally lensed by foreground clusters [36] [37]. These arcs are formed when a galaxy is nearly imaged into an Einstein ring. Measurements of the cluster and background galaxy redshifts yield an estimate of the cluster mass within the impact parameter of the lens; for most of the cases studied, these estimates are in reasonable agreement with the mass–to–light ratios inferred from the virial theorem, but it should be noted that the arc observations probe only the inner few hundred kpc of the clusters. (Work is presently being done to extend this idea to map clusters on larger scales, by looking for more minute statistical distortions in the alignments of background galaxies.) Typical inferred cluster values are $\Upsilon \sim 100 - 250h$, which would imply $\Omega_0 \sim 0.1 - 0.2$. Since clusters are rare objects, occupying a very small fraction of the universe, one expects this to be a lower bound, in which case some form of *non-baryonic* dark matter is probably required, given the limits on $\Omega_B$ above.

X-ray observations, most recently by the ROSAT satellite,[38] have begun to map the density and temperature profiles of the hot gas which permeates many clusters. Since the gas is in hydrostatic equilibrium, it can be used to trace the cluster mass distribution (including dark matter) directly,

$$M_{tot}(r) = -\frac{kT(r)}{G\mu m_p}\left(\frac{d\ln n}{d\ln r} + \frac{d\ln T}{d\ln r}\right) \tag{4.4}$$

where $\mu$ is the mean molecular weight of the gas, and $n(r)$ and $T(r)$ are the gas density and temperature. Cluster masses inferred from X-ray observations are generally comparable to but $\sim 30\%$ less than virial estimates. Consistent with this, the X-ray measurements indicate that clusters are surprisingly baryon-rich, in the sense that the gas constitutes typically $5 - 10h^{-3/2}\%$ of the inferred binding mass within approximately $1h^{-1}$ Mpc of the center of a rich cluster like Coma. (Moreover, the dark matter tends to be *more* centrally concentrated than the gas out to this scale, suggesting that the gas fraction of the whole cluster is at least as large as the value above.[40]) If this ratio is representative of the baryon mass fraction of the universe, then the nucleosynthesis bound on $\Omega_B$ would imply[41] the *upper* limit $\Omega_0 \lesssim 0.15h^{-1/2}$. (Including the baryon stellar component in cluster galaxies only strengthens this limit.) This has been taken as evidence against the universe having closure density ($\Omega_0 = 1$) and would require advocates of inflation (which implies $k = 0$) to fall back on a cosmological constant or some other smoothly distributed component. The other possibility would be to loosen the nucleosynthesis bounds on $\Omega_B$ through some non-standard scenario such as inhomogeneous nucleosynthesis, but the upper bound on $\Omega_B$ is not raised sufficiently in this model to get around the argument above. One should perhaps be cautious about inferring the universal baryon fraction from a rich cluster like Coma. In particular, clusters with higher X-ray temperatures (and therefore larger total masses) may have a larger fraction of their total mass in gas.[40] This trend, coupled with the steeply falling distribution function of cluster temperatures, $dn_c/dT \sim T^{-5}$ for $3 < kT < 10$ keV, suggests that the mean gas (and baryon) fraction may be substantially below the value for Coma, since the mean is dominated by the more numerous cooler clusters ($kT_{Coma} \simeq 7$ keV). This would raise the derived upper bound on $\Omega_0$ closer to unity and suggests that massive clusters like Coma may not be representative of the baryon fraction of the Universe. On the other hand, N-body simulations of cold dark matter with baryons[41] indicate that the baryon fraction in a Coma-size cluster should be representative of the mean, so the present situation is somewhat confusing.

### 4.3. Large-scale flows

Moving to still larger scales, the deviations from the Hubble flow have been used to infer the cosmic density over scales of order $50h^{-1}$ Mpc. The basic idea is to compare samples of the density perturbation field and the peculiar velocity field covering the same volume;[42] assuming they arise gravitationally, the proportionality between them depends on the rate of growth of the density fluctuations, which in turn depends on $\Omega_0$. On very large scales, as discussed above, the rms density fluctuations are small, so linear perturbation theory away from the FRW spacetime is a reasonable first approximation (see the lectures by Scherrer). In this case one finds the relation

$$\nabla \cdot \vec{v}_p = -H_0 \Omega_0^{0.6} \delta \quad , \qquad (4.5)$$

where the density field $\delta(\vec{x}) = (\rho(\vec{x}) - \bar{\rho})/\bar{\rho}$, and the perturbation growth rate enters through $d\ln\delta/d\ln a \simeq \Omega^{0.6}$. If one expresses distances $d$ in terms of their equivalent Hubble velocities, $v = H_0 d$, then $H_0$ drops out of Eqn.(4.5), so the uncertainty in the Hubble parameter does not undermine this method. A number of different approaches have been used to extract $\Omega_0$ in this way, and they have all given consistently high answers, $\Omega_0 \gtrsim 0.5$. The topographic correlation between the observed galaxy field and the density field inferred from galaxy peculiar velocity surveys suggests that galaxies do broadly trace the mass distribution on large scales, in the sense that more galaxies are found in regions of high mass density (that is, high $\nabla \cdot \vec{v}_p$), but the galaxy distribution may be 'biased' with respect to the mass. In the simplest linear bias model, the smoothed galaxy and mass density fields are assumed to be proportional, $\delta_{gal}(\vec{x}) = b_{gal} \delta(\vec{x})$, where $b_{gal}$ is the bias factor, taken to be constant for a given class of galaxies. Since $\delta_{gal}$ is what is observed, the comparisons based on Eqn.(4.5) actually constrain the combination $\Omega_0^{0.6}/b_{gal}$, where the bias factor refers, e.g., to galaxies selected from the IRAS catalog. Recent determinations have found $\Omega_0^{0.6}/b_{gal} \sim 1$. For a bias factor of order unity, this is consistent with $\Omega_0 = 1$, which is pleasing to theorists and also buttresses the case for non-baryonic dark matter. It is well to keep in mind, however, that biasing is presumably a complex process associated with the non-linear stages of galaxy formation, so that the proportionality of the galaxy and density fields may be non-linear and/or scale-dependent. In addition, even with substantial smoothing of the density field, the perturbation amplitude in many regions is not small, and corrections to the linear theory must be taken into account.

A summary of typical estimates of the density parameter on different scales is given in Table 1.

### 4.4. Non-baryonic Dark Matter

If $\Omega_0 > \Omega_B$, as the dynamical observations on scales larger than clusters suggest, some form of non-baryonic matter must be invoked to make up the balance. Moreover, since this matter must be dark ($\Omega_0 \gg \Omega_{lum}$), it is natural to consider weakly interacting particles as candidates. It is convenient to distinguish two broad classes of non-baryonic dark matter, hot and cold, on the basis of their clustering properties. The prototypical hot dark matter candidate is a light neutrino with mass $m_\nu \simeq 20$ eV (see Sec. 6.3). Since they are relativistic ($m_\nu \lesssim T$) until relatively recent epochs, light neutrinos would free-stream out of and damp out density perturbations up to the scale of galaxy clusters. Galaxies would form after clusters via fragmentation ('top down'). Due to phase-space constraints[43], light neutrinos would not cluster significantly on the scale of galaxies: baryons would constitute the predominant dark matter in galaxy halos, while neutrinos would dominate in clusters. Cold dark matter, on the other hand, is defined to have negligible free-streaming length–it clusters on all scales. In cold dark matter models, structure generally forms hierarchically, with smaller clumps merging to form larger ones ('bottom up'). For this reason, many (but not all) theorists since the early '80's have tended to prefer cold over hot dark matter, but it is worth noting the recently surging popularity of a mix 'n match scenario: a combination of 70% cold and 30% hot dark matter (say, with $m_\nu \simeq 7$ eV) produces a favorable spectrum of large-scale density perturbations in the context of inflation, with some apparent advantages over pure cold dark matter.

Table 1: Estimates of the cosmic density parameter $\Omega_0$.

| System | $\Upsilon$ | $\Omega_0$ |
|---|---|---|
| Luminous matter in galaxies | 5 - 12h | 0.007 |
| Galaxy halos | 35 | $0.02h^{-1}$ |
| Clusters | 250h | 0.2 |
| Large-scale flows | – | 0.5 -1 |

The theoretically favorite candidates for cold dark matter are weakly interacting massive particles (WIMPs), with masses generally in the range $20-150$ GeV, and the axion, an ultra-light pseudoscalar with a mass of order $10^{-5}$ eV. The most attractive WIMP candidate is the neutralino, the lightest supersymmetric fermionic partner of the standard model bosons; its weak annihilation rate in the early universe naturally leaves it with an abundance comparable to the present critical density. The axion is the pseudo-Nambu-Goldstone boson associated with spontaneous breakdown of a global $U(1)$ symmetry (the Peccei-Quinn symmetry) introduced to explain why the strong interactions conserve CP. The global symmetry is spontaneously broken at some large mass scale $f_{PQ}$, through the vacuum expectation value of a complex scalar field, $\langle \Phi \rangle = f_{PQ} \exp(ia/f_{PQ})/\sqrt{2}$. At energies below the scale $f_{PQ}$, the only relevant degree of freedom is the massless axion field $a$, the angular mode around the bottom of the $\Phi$ potential. At a much lower energy scale, $\Lambda_{QCD} \sim 100$ MeV, the symmetry is *explicitly* broken when QCD becomes strong, and the axion obtains a periodic potential of height $\sim \Lambda_{QCD}^4$. In 'invisible' axion models with Peccei-Quinn scale $f_{PQ} \sim 10^{12}$ GeV, the resulting axion mass is $m_a \sim \Lambda_{QCD}^2/f_{PQ} \sim 10^{-5}$ eV and $\Omega_a \sim 1$. Although light, invisible axions interact so weakly, with cross-section $\sigma \sim 1/f_{PQ}^2$, that they were never in thermal equilibrium: they form as a cold Bose condensate.

Accelerator searches for neutrino mixing and beta-decay experiments on neutrino mass should provide useful constraints on the possibility of neutrino dark matter. Active experimental efforts are also underway to detect both WIMPs and axions. Direct WIMP detection looks for the signals produced when a halo WIMP collides with a nucleus in a kg-size cryogenic crystal, depositing of order 10 keV in ionization and phonons (detector schemes based on scintillation and excitations of superfluids and superconductors are also being developed). Indirect WIMP detectors search for high energy neutrinos produced when WIMPs annihilate in the Sun and the Earth; large underground or underwater detectors currently in place or under development with sensitivity to WIMP annihilations include Kamiokande, MACRO, AMANDA, and DUMAND. Accelerator searches for supersymmetry also will constrain the neutralino parameter space. A large-scale axion search based at Livermore, expected to come on-line in 1994, will search for resonant conversion of halo axions to microwave photons in a cavity with a strong magnetic field. A scaled-up version of an idea originally proposed by Sikivie, this detector should approach the cosmologically interesting region of axion couplings for the first time[44].

Finally, given the paucity of direct evidence for dark matter, it is probably healthy to keep an open mind to alternatives. While it is natural to ascribe flat galaxy rotation curves and large cluster velocity dispersions to unseen matter, Milgrom[45] and others have argued that they may instead signal a breakdown of Newton's law of inertia at very low acceleration. The extent to which Milgrom's modified Newtonian dynamics (MOND) accounts for all the phenomena normally imputed to dark matter is controversial, and a full theory with which one could explore cosmology has been lacking. Nevertheless, at a minimum it provides a useful challenge to the accepted dogma. On a relatively more conservative side, the possibility that the dark matter interacts by other long-range forces in addition to gravity has recently been explored.[63] Such interactions are significantly constrained by galaxy and cluster observations, but could nevertheless have interesting implications for structure formation and biasing.

## 5. Cosmological Parameters: Taking Stock

With this brief survey in hand, it is useful to pause and place these numbers for the cosmological parameters in theoretical context. If an extended period of inflation took place in the early universe (see below), then the spatial geometry should now be observationally indistinguishable from $k = 0$. If the cosmological constant vanishes, from Eqn.(3.2) spatial flatness implies $\Omega_0 = 1$ (with the concomitant requirement of non-baryonic dark matter), and thus $t_0 = 2/3H_0 = 6.5h^{-1} \times 10^9$ yr. This is uncomfortably low compared to globular cluster ages unless $h \lesssim 0.5$ ($t_0 \gtrsim 13$ Gyr) and definitely problematic unless $h < 0.65$ ($t_0 > 10$ Gyr), still on the low side of the $H_0$ observations. However, a non-vanishing $\lambda_0$ is certainly allowed at some level by observations and has sporadically come into vogue, most recently to alleviate both this age problem and the large-scale power problem for inflationary cosmology. For example, in a flat model with $\Omega_0 = 0.25$, $\lambda_0 = 0.75$, Eqn.(3.4) gives $t_0 \simeq 1/H_0 = 9.75h^{-1}$ Gyr, consistent with the lower bound $t_{gc} > 10$ Gyr for the entire observed range of $H_0$, and yielding a healthy, $t_0 > 13$ Gyr-old universe for $h < 0.75$. This lower value of $\Omega_0$ is consistent with the dynamical mass estimates from clusters, but somewhat below the recent estimates from large-scale flows. Since there is currently little theoretical guidance as to why $|\lambda_0|$ is as small as it is, and no firm proof that it should vanish, it is probably best to keep an open mind, although the fact that we would be living just at the epoch when $\Omega_0$ is comparable to $\lambda_0$ might seem to beg for explanation. The third possibility is that theoretical prejudice is wrong, and that we live in an open, low-density, perhaps purely baryonic universe with negligible $\Lambda$, in which case the globular cluster age range is also compatible with somewhat larger values of $H_0$. The challenge in this case is to explain the large-scale flows (or attribute them to systematic distance errors) and to form large-scale structure without violating CMBR anisotropy constraints. In this case, it also would be an odd coincidence that we live just at the epoch when the curvature term in (3.2) is becoming appreciable compared to the matter term.

## 6. The Early Universe

We have seen that the FRW models provide a natural framework for understanding observations of the universe on the largest scales, in particular the Hubble law. Aside from the Hubble expansion, two other observational pillars support the temple of the standard cosmology: the thermal spectrum of the CMBR and the light element abundances. Both direct our attention backward to much earlier epochs in the history of the universe: the hot big bang.

### 6.1. The Cosmic Microwave Background

The 2.7 K cosmic microwave background radiation, discovered thirty years ago, has since been shown to be remarkably isotropic (to roughly one part in $10^5$ on angular scales larger than a few arc minutes) and to have an exactly thermal blackbody spectrum to within the experimental errors. The spectral measurements cover the range of photon wavelengths from roughly 0.05 - 50 cm, while the most impressive verification of the Planck spectrum comes from the recent COBE FIRAS observations in the range 0.05 - 1 cm. The measured FIRAS flux vs. frequency[47] is shown in Fig. 6; it is indistinguishable from a Planck spectrum with a temperature of $T = 2.726 \pm 0.010$ K. The error bars on the points are too small to see, so they have been multiplied by a factor of 100 in the figure.

The blackbody nature of the CMBR is indicative of a medium that has relaxed to thermal equilibrium. However, there are strong grounds for believing that the recent universe is transparent to CMBR photons, that is, that the mean free path for microwave photon scattering is much longer than the present Hubble length: (i) we observe radio galaxies as point sources out to redshifts of order $z \sim 4$, which requires microwave transparency out to this scale; (ii) even postulating a fully ionized intergalactic medium with closure density ($\Omega_{igm} \sim 1$), the mean free time for Compton scattering is longer than the Hubble time back to a redshift $z \gtrsim 10$. Therefore, the CMBR is not now in thermodynamic equilibrium and could not have been thermalized at recent cosmological epochs. The standard explanation is that the CMBR is a remnant of a much earlier hot, dense phase of the universe when the matter and radiation were in equilibrium. Indeed, running the Hubble expansion backward in time, this is just what one would expect: as the matter is compressed, it heats up, becoming a dense, ionized plasma which efficiently scatters and thermalizes the radiation on a timescale much shorter than the expansion timescale.

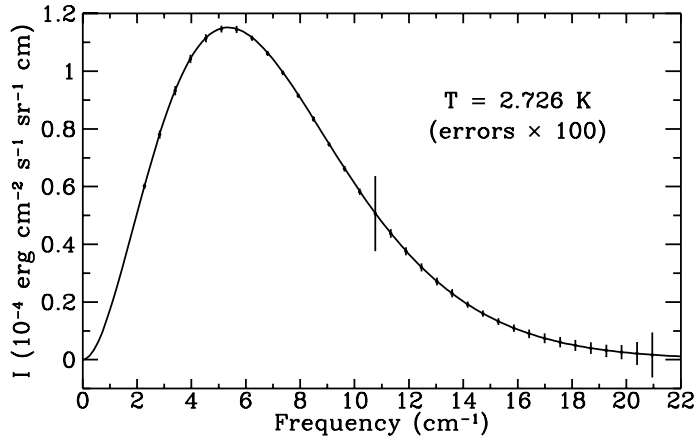

Figure 6: CMBR brightness vs. frequency (inverse wavelength) measured by COBE FIRAS.

This argument suggests that the early universe can be accurately described as a dilute, adiabatically expanding gas of particles and radiation in local thermal equilibrium, uniquely characterized by the instantaneous temperature $T(t)$. While this may reflect conditions over much of the early history of the universe, we know it is not the full story, because a gas in thermal equilibrium is featureless and structureless, and the observed universe is not so boring. Although thermal equilibrium is a good starting point, the interesting epochs in cosmic history–the ones which leave potentially observable signatures or relics, such as the light elements, the baryon asymmetry, and particle dark matter–are those in which a particle species $i$ goes out of thermal equilibrium, because the rate of interactions keeping it in equilibrium, $\Gamma_i$, falls below the expansion rate $H(t)$. This happens at a temperature $T_F$ defined by

$$\left(\frac{\Gamma}{H}\right)_{T_F} \simeq 1 \ . \qquad (6.1)$$

$T_F$ is called the freeze-out temperature: at temperatures $T < T_F$, the particle's comoving number density is fixed, $n_i \sim a^{-3}$. Thus, the study of the early universe requires both equilibrium and non-equilibrium thermodynamics (the Boltzmann equation). For completeness, it is worth noting that there may also be particle relics, such as magnetic monopoles and axions, which were never in thermal equilibrium.

It is helpful to have some characteristic numbers for the CMBR. For ultra-relativistic particles in kinetic equilibrium at temperature $T$, the energy density is

$$\rho_i = \frac{\pi^2}{30} g_i T^4 \times \begin{cases} 1 & \text{bosons} \\ 7/8 & \text{fermions} \end{cases} \tag{6.2}$$

and the number density

$$n_i = \frac{\zeta(3)}{\pi^2} g_i T^3 \times \begin{cases} 1 & \text{bosons} \\ 3/4 & \text{fermions} \end{cases} \tag{6.3}$$

where $g_i$ is the number of internal degrees of freedom, and the Riemann zeta function $\zeta(3) \simeq 1.202$. For blackbody radiation at $T_0 = 2.73$ K, the present CMBR energy density is $\rho_\gamma = 4.8 \times 10^{-34}$ gm/cm$^3$, and the photon number density is $n_\gamma = 420$ cm$^{-3}$. For comparison, the present baryon density is $n_B = \rho_B/m_p = 1.12 \times 10^{-5} \Omega_B h^2$ cm$^{-3}$, and the baryon to photon ratio is

$$\eta \equiv \frac{n_B}{n_\gamma} = 2.7 \times 10^8 \Omega_B h^2 \quad. \tag{6.4}$$

In the early universe, baryons and photons are kept in thermal contact by Compton scattering, $e^- \gamma \to e^- \gamma$. As the universe expands and cools, eventually it becomes energetically favorable for the ionized plasma to recombine to form neutral hydrogen, $e^- + p \to H + \gamma$. When this happens, the density of free electrons drops precipitously, and the Compton scattering rate per photon, $\gamma_\gamma = n_e \sigma_{Comp}$, falls below the expansion rate $H$; as a result, the photons decouple (freeze out) from the matter. In the absence of subsequent reionization of the matter, the photons will have travelled freely since that epoch, which is therefore called the *surface of last scattering*.

It is a useful exercise to estimate when hydrogen recombination and photon decoupling took place. Since the recombination rate is initially rapid compared to the expansion rate, the baryons and photons maintain ionization equilibrium. The densities of the non-relativistic baryons in thermal equilibrium are

$$n_i = g_i \left(\frac{m_i T}{2\pi}\right)^{3/2} \exp\left(\frac{\mu_i - m_i}{T}\right) \quad, \tag{6.5}$$

where $i = e, p, H$, and $\mu_i$ are chemical potentials. Defining the ionization fraction, $x = n_p/(n_p + n_H)$, and using chemical equilibrium to relate the chemical potentials, $\mu_p + \mu_e = \mu_H$, we arrive at the Saha equation for the equilibrium ionization fraction,

$$\frac{1 - x_{eq}}{x_{eq}^2} = \frac{4\sqrt{2}\zeta(3)}{\sqrt{\pi}} \eta \left(\frac{T}{m_e}\right)^{3/2} \exp\left(\frac{B}{T}\right) \quad, \tag{6.6}$$

where $B = m_e + m_p - m_H = 13.6$ eV is the binding energy of neutral hydrogen. Using (6.4) for $\eta$ and taking $\Omega_B h^2 \simeq 0.01$ from nucleosynthesis, one finds that $x_{eq}$ drops rapidly away from unity–that is, the hydrogen effectively recombines– at a temperature $T_{rec} \simeq 0.3$ eV. Since the temperature scales as $T = 2.73(1+z)$ K, this corresponds to the epoch $z_{rec} \simeq 1300$. This result is just what one expects: recombination takes place when there are too few photons left in the high-energy tail of the thermal distribution to ionize the hydrogen. This happens at a temperature somewhat below the binding energy of hydrogen, because the number of photons per baryon is so large ($\eta \ll 1$). Once $x_{eq}$ drops appreciably below unity, the Compton rate per photon, $\Gamma_\gamma = x \eta n_\gamma \sigma_{Comp}$ soon falls below the expansion rate, and the photons decouple at a redshift $z_{dec} \simeq 1100$. From (2.19), this corresponds to a decoupling time $t_{dec} = 5.6 \times 10^{12} h^{-1}$ sec $\simeq 180,000 h^{-1}$ yr (for $\Omega_0 = 1$).

If photons decoupled at such an early epoch, why does the CMBR still have such a precisely thermal spectrum? The simple reason is that the expansion preserves the blackbody thermal spectrum even in the absence of thermalizing interactions. At temperatures $T > T_{dec}$, the photons are in kinetic equilibrium, with a Bose distribution function,

$$f(\mathbf{p}, t < t_{dec}) = \frac{1}{\exp(E/T) - 1} \quad , \tag{6.7}$$

where $E = |\mathbf{p}|$. After decoupling, the photons are free particles, their energies just redshifting with the expansion, $E(t) = E(t_{dec}) a(t_{dec})/a(t)$. However, since they are no longer interacting, the occupation number must be conserved for each momentum state. As a result, the distribution function maintains the form (6.7), with effective temperature given by $T(t) = T(t_{dec})(a(t_{dec})/a(t))$; since this holds for each momentum state, the spectrum remains thermal.

In fact, the issue of the preservation of the thermal spectrum is more subtle than this argument suggests, and the spectrum observation by FIRAS a more powerful tool[48], because we have not taken into account the possibility that some physical process could distort the spectrum away from the form (6.7). For example, annihilating or decaying elementary particles or exploding black holes could release energy into the photon-baryon plasma that is not completely thermalized by the time of photon decoupling. Alternatively, after decoupling, an early generation of stars, quasars, or active galactic nuclei might release enough energy to heat up and reionize the baryons, which would then scatter off the CMBR and distort the spectrum. The kind of spectral distortion produced depends on the epoch when such a process releases energy into the photon-baryon system. Energy released into the baryons at $z_{rel} \gtrsim 3 \times 10^6$ is completely thermalized by Compton scattering and bremmstrahlung emission: the photons relax to a new equilibrium distribution at a higher temperature, leaving no distortion. Energy released in the interval $10^5 \lesssim z_{rel} \lesssim 3 \times 10^6$ cannot be completely thermalized, because photon production becomes inefficient; in this regime, multiple Compton scattering leads to a Bose-Einstein photon distribution,

$$f(p) = \frac{1}{\exp\left(\frac{E-\mu}{T}\right) - 1} \quad , \tag{6.8}$$

where the chemical potential $\mu$ is related to the energy density injected, $\Delta\rho$, by $\Delta\rho/\rho_\gamma = 0.7\mu/T$. The FIRAS results shown in Fig. 6 can be translated into the bound $|\mu/T| < 3.3 \times 10^{-4}$, severely restricting the energy density injected at such times. If energy is released into the electrons later, at $z_{dec} = 10^3 < z_{rel} < 10^5$, $e^-\gamma$ scattering produces a Compton $y$-distortion which can be expressed as a frequency-dependent temperature,

$$\frac{\delta T(\nu)}{T} \simeq y\left(\frac{\nu}{T}\frac{e^{\nu/T}+1}{e^{\nu/T}-1} - 4\right) \to \begin{cases} -2y & \text{for } \nu/T \ll 1 \\ y\nu/T & \text{for } \nu/T \gg 1 \end{cases} . \tag{6.9}$$

The $y$-parameter is related to the energy release by $y = \Delta\rho/4\rho_\gamma$. Eqn.(6.9) states that Compton scattering by hot electrons at $T_e > T_\gamma$ moves photons from low to high frequency, depleting the spectrum in the low-frequency Rayleigh-Jeans region and enhancing it in the Wien part of the spectrum. Comparison with the data in Fig.6 yields the observational constraint $y < 2.5 \times 10^{-5}$. These bounds together severely constrain any model in which energy is released between roughly one and 300,000 yr after the big bang.

## 6.2. The Radiation Era

Summing eqn.(6.2) over all effectively massless species in equilibrium, we can define the total radiation energy density by

$$\rho_r = \frac{\pi^2}{30}g_*T^4 \ , \tag{6.10}$$

where $g_*(T)$ counts the number of effective relativistic Bose (B) and Fermi (F) degrees of freedom, which may have different effective temperatures,

$$g_*(T) = \sum_{i=B} g_i\left(\frac{T_i}{T}\right)^4 + \frac{7}{8}\sum_{i=F} g_i\left(\frac{T_i}{T}\right)^4 \ . \tag{6.11}$$

As the universe expands and cools, the function $g_*(T)$ decreases whenever $T$ drops below the mass $m_i$ of a particle species. For the photon and assuming 3 massless neutrinos, it is currently $g_*(T_0) = 3.36$; above the $e^+e^-$ mass threshold, $g_*(1 - 100\text{MeV}) = 10.75$; and at temperatures above the masses of all particles in the standard electroweak model, $g_*(T > 300\text{GeV}) \gtrsim 100$.

From (6.11), and assuming there are no exotic heretofore unknown relativistic particles, the present radiation energy density is $\rho_r(t_0) = 8.1 \times 10^{-34}$ gm/cm$^3$. Comparing with the critical density in eqn.(4.1), the radiation makes a negligible contribution to the present density,

$$\Omega_r = 4.3 \times 10^{-5}h^{-2} \ , \tag{6.12}$$

confirming our assumption in Sec. 2 that the present universe is matter-dominated. On the other hand, since $\rho_r/\rho_m \sim a^{-1}$, the universe was radiation-dominated at early times. From (6.12), the transition from radiation to matter-domination ($\rho_m = \rho_r$) occurred when

$$1 + z_{eq} = \frac{a_0}{a_{eq}} = \frac{T_{eq}}{T_0} = 2.3 \times 10^4 \Omega_0 h^2 \quad , \quad (6.13)$$

which corresponds to a photon temperature $T_{eq} = 5.5\Omega_0 h^2$ eV. For $\Omega_0 = 1$, this happened when the age of the universe was $t_{eq} = 4.4 \times 10^{10} h^{-4}$ sec. Note that for $\Omega_0 = 1 \gg \Omega_B$, the universe becomes matter-dominated before (but not long before) recombination.

During the radiation-dominated (RD) epoch ($t < t_{eq}$), the spatial curvature term in the Friedmann equation can be neglected, since it scales with $a$ more slowly than $\rho_m$ or $\rho_r$ and we know that it is at most comparable to the matter density term today. The RD era solution is therefore described by (2.20). Combining (2.20) and (6.10), we can relate the temperature and expansion rate in the RD epoch:

$$H = 1.66 g_*^{1/2} \frac{T^2}{M_{Pl}} \quad . \quad (6.14)$$

Since $H = 1/2t$ from (2.20), we can also relate temperature and age:

$$t = \frac{0.3 M_{Pl}}{g_*^{1/2} T^2} = \frac{2.4}{g_*^{1/2}} \left(\frac{T}{\text{MeV}}\right)^{-2} \text{ sec} \quad . \quad (6.15)$$

Table 2: Thermal history of the universe.

| $T$(GeV) | $t$(sec) | Event |
|---|---|---|
| $10^{19}$ | $10^{-43}$ | Planck era: quantum gravity |
| $10^{15}$ | $10^{-35}$ | Grand unification: inflation, topological defects |
| $10^3$ | $10^{-12}$ | Supersymmetry, technicolor |
| $10^2$ | $10^{-10}$ | Electroweak transition: $SU(2) \times U(1) \to U(1)_{em}$ |
| 0.2 | $10^{-4.5}$ | Quark-hadron transition; chiral symmetry breaking |
| $10^{-3}$ | 1 | Neutrino decoupling |
| $5 \times 10^{-4}$ | 4 | $e^+e^-$ annihilation |
| $7 \times 10^{-5}$ | 200 | Big Bang nucleosynthesis |
| $10^{-9}$ | $5 \times 10^{11}$ | Matter-radiation equality |
| $3 \times 10^{-10}$ | $10^{13}$ | $H$ recombination; photon decoupling |
| $6 \times 10^{-13}$ | $10^{17}$ | Non-linear structures form |
| $3 \times 10^{-13}$ | $4 \times 10^{17}$ | Chicago Bulls three-peat |
| | | (note: this was given as a *prediction* at TASI '93) |

A timeline for the early universe with some important epochs delineated is shown in Table 2. The CMBR probes conditions back to the time of photon decoupling, while the light element abundances probe conditions at temperatures comparable to nuclear binding energies, $T \sim$ MeV. Going backward in time, QCD predicts that chiral symmetry should be broken at a temperature of order 100 MeV. At about that time, quarks should also become confined inside hadrons, but it is not clear if this is a smooth (second-order) transition or a first-order transition involving the release of latent heat and the nucleation of hadron bubbles. If the quark-hadron transition is first-order, the resulting inhomogeneous baryon distribution might leave observable signatures in the light element abundances.

Going to earlier times, the electroweak symmetry should be restored at a temperature of order 100 GeV. The dynamics of this transition is currently an area of active investigation: if it is a first-order transition, it is possible that the resulting non-equilibrium conditions were ripe for the generation of the baryon asymmetry. If baryons and anti-baryons had been present in exactly equal numbers in the early universe, their annihilation would eventually have driven the baryon to photon ratio down to a value many orders of magnitude smaller than (6.4). Thus, when the baryons were relativistic, there must have been a small asymmetry between the density of baryons and anti-baryons, $n_B/n_b = (n_b - n_{\bar{b}})/n_b \sim \eta$. As first pointed out by Sakharov, the generation of such a baryon asymmetry requires baryon-number and CP-violating interactions as well as a departure from thermal equilibrium (since $n_b = n_{\bar{b}}$ in equilibrium). It is currently thought that baryogenesis takes place at the GUT or electroweak eras.

Going back earlier than $10^{-10}$ sec, we must invoke physics beyond the standard electroweak model, and the events become increasingly speculative. Particle physics models suggest there may be new physics lurking at the TeV scale–perhaps supersymmetry, technicolor, or various extensions of the standard model. In the simplest grand unified theories, the strong and electroweak interactions are unified (the symmetry between them restored) at an energy scale of order $10^{15}$ GeV. A cosmological phase transition at that epoch might lead to inflation or to the generation of topological defects such as monopoles, cosmic strings, or textures; the resulting density and gravitational wave perturbations produced could provide the seeds for large-scale structure and leave a signature in the CMBR anisotropy. Classical cosmology runs into a wall at the Planck era, $t \sim 10^{-43}$ sec: at that epoch, quantum fluctuations in the spacetime metric are expected to be large, and a quantum theory of gravity is required. If superstrings provide the fundamental description of nature, inherently stringy effects would become important around that scale.

### 6.3. Relic Neutrinos and Hot Dark Matter

As an interesting example of interactions freezing out and leaving a relic species, consider light neutrinos. The cross-section for $\nu_e e^- \to \nu_e e^-$ scattering through $W$ and $Z$ exchange is of order $\sigma_\nu \sim \alpha_w^2 p^2/(p^2-M^2)^2$, where $\alpha_w$ is the weak fine structure constant, $M \simeq M_Z, M_W$ is the gauge boson mass, and $p$ are the fermion momenta in the center of mass. First consider the very early universe, at temperatures $T \gg M$. In this case, the momenta $p \sim T \gg M$, resulting in a cross-section $\langle \sigma v \rangle \sim \alpha_w^2/T^2$.

Since the particle density $n \sim T^3$, the interaction rate is $\Gamma_\nu \sim \alpha_w^2 T$. Comparing with the expansion rate, $H \sim T^2/M_{Pl}$, we find that the interaction rate is larger, $\Gamma_w/H > 1$, at temperatures $T \lesssim \alpha_w^2 M_{Pl} \sim 10^{16}$ GeV. If we begin our consideration of cosmology at the Planck era, $T \sim 10^{19}$ GeV, this tells us that gauge interactions are initially out of equilibrium, but local thermal equilibrium is a good working assumption below about $10^{16}$ GeV. As a result, inferences about physics near the Planck scale which rely on equilibrium thermodynamics should be viewed with skepticism.

The situation changes when the temperature falls below the gauge boson masses, $T \ll M$. In this regime, we have $M \gg T \sim p \gg m_e$, and the $\nu e^-$ cross-section becomes $\langle \sigma v \rangle \sim \alpha_w^2 T^2/M^4 \sim G_F^2 T^2$, where $G_F$ is the Fermi constant, $G_F \simeq 10^{-5}$ GeV$^{-2}$. Comparison with the expansion rate now gives $\Gamma_w/H \sim G_F^2 M_{Pl} T^3$, which falls below unity at the neutrino freeze-out temperature, $T_F \simeq 1$ MeV. For light neutrinos, we have $T_F \gg m_\nu$, so the neutrinos are ultrarelativistic, with a density comparable to photons, when they freeze out; such particles are called *hot* relics, as opposed to *cold* relics, which are non-relativistic ($T_F \lesssim m$) when they freeze out.

Consider the implications of neutrino decoupling for the present neutrino density. We will make use of the fact that, in thermal equilibrium, the total entropy in a comoving volume is conserved by the first law of thermodynamics, $S = sa^3 =$ constant, where the entropy density is

$$s = \frac{p + \rho}{T} \ . \tag{6.16}$$

Soon after the neutrinos decouple, the electrons and positrons annihilate, $e^+ e^- \to \gamma\gamma$, at $T \sim m_e \sim 0.5$ MeV, converting their entropy into photons. Since this takes place after $\nu$ freeze out, the neutrinos do not partake of the entropy boost. Before annihilation, the $e^+ e^-$ entropy is $s_e = 4 \times (7/8) \times (2\pi^2/45) T^3$, and the photon entropy $s_\gamma = 2 \times (2\pi^2/45) T^3$. After annihilation, the photon entropy has been increased by the factor

$$\frac{s'_\gamma}{s_\gamma} = \frac{2 + 7/2}{2} = \frac{11}{4} \ , \tag{6.17}$$

which means the photons are heated with respect to the neutrinos,

$$\frac{T'_\gamma}{T_\nu} = \left(\frac{s'_\gamma}{s_\gamma}\right)^{1/3} = 1.4 \ . \tag{6.18}$$

Using the present CMBR temperature from COBE, this implies a present neutrino temperature of $T_\nu = 1.95$ K, and a neutrino density (for a single neutrino species)

$$n_\nu = (3/4)(T_\nu/T_\gamma)^3 n_\gamma = (3/11) n_\gamma = 115 \text{ cm}^{-3} \ . \tag{6.19}$$

We can also use this to calculate the present energy density in radiation: assuming all three neutrino species are massless, we have

$$g_*(T_0) = 2 + 3 \times 2 \times \frac{7}{8} \left(\frac{4}{11}\right)^{4/3} = 3.36 \ , \tag{6.20}$$

which leads directly to (6.12).

It is possible that one or more of the neutrinos is massive, the current experimental neutrino mass limits being $m_{\nu_e} < 9$ eV, $m_{\nu_\mu} < 270$ keV, $m_{\nu_\tau} < 35$ MeV. If a neutrino is massive and stable (or long-lived compared to the age of the universe), it could contribute more substantially to the density of the universe. From (6.19) and (4.1), one finds

$$\Omega_\nu h^2 = \frac{m_\nu n_\nu}{\rho_c h^{-2}} = \frac{m_\nu}{92} \text{ eV} \quad . \tag{6.21}$$

From sec. 5, neutrinos could provide the dark matter and close the universe, $\Omega_\nu = 1$, if $h \simeq 0.5$, which would require a mass $m_\nu \simeq 20$ eV. While larger than the $\nu_e$ mass limit, either the muon and tau neutrinos could have masses in this range; moreover, even an electron neutrino at the upper end of the experimental mass limit would contribute substantially to the energy density.

If neutrinos are the dark matter, an interesting argument due to Tremaine and Gunn[43] shows that they cannot cluster in galaxies, so they would not provide the dark halos of the kind shown in Fig. 5. To see the point, suppose neutrinos were clustered in the halo of a galaxy like the Milky Way. Their number density would be $n_\nu = \int d^3 p f_\nu(p) \sim f_\nu \langle p^3 \rangle \sim f_\nu (m_\nu v_h)^3$, where the halo velocity dispersion is $v_h \sim 200$ km/sec. By the Pauli exclusion principle, the neutrino occupation number must satisfy $f_\nu(p) \leq 1$, which implies the constraint $n_\nu/(m_\nu v_h)^3 = \rho_\nu/(m_\nu^4 v_h^3) \leq 1$. The local halo density in the model of Fig. 5 is of order $\rho_h \sim 0.4$ GeV cm$^{-3}$; for this to be composed of neutrinos would require $m_\nu \geq (\rho_h/v_h^3)^{1/4} \simeq 50$ eV. Extending this argument to dwarf galaxies leads to the tighter constraint $m_\nu \gtrsim$ (100 - several hundred) eV. These numbers are significantly higher than the closure density neutrino mass; turning the argument around, we conclude that in a neutrino-dominated universe, galaxy halos must be baryonic.

### 6.4. Relic WIMPS and Cold Dark Matter

In the discussion above, we assumed that the neutrinos are light compared to their freeze-out temperature, $m_\nu \ll T_F \sim 1$ MeV. If neutrinos, or some other weakly interacting particle, were much heavier, the situation changes: the particle becomes non-relativistic before freeze-out, so its abundance relative to radiation is depleted by annihilations–such particles are collectively known as cold relics or WIMPs. Examples of cold relics are supersymmetric neutralinos, fermionic partners of standard model bosons, with masses in the range $m \sim 10$ GeV $-1$ TeV. If R-parity is conserved, the lightest supersymmetric particle (LSP) is stable and is a prime candidate for (cold) dark matter with $\Omega_{LSP} \sim 1$.

Accurate calculations of relic WIMP abundances generally require numerical solution of the Boltzmann equation, which describes how the WIMP abundance relative to its equilibrium value evolves over time in the expanding universe. However, one can get an order of magnitude estimate from eqn.(6.1), which states that the relic particle abundance (relative to the entropy density) is approximately the equilibrium abundance at freeze-out. Using the annihilation rate at equilibrium, $\Gamma = n_{eq}\langle \sigma v \rangle$, with $n_{eq}$ given by the Boltzmann form (6.5) (here assuming $\mu = 0$) and the expansion rate in (6.14), the freeze-out condition becomes (dropping factors of $2\pi$)

$$\left(\frac{\Gamma}{H}\right)_{T_F} \simeq \frac{(mT_F)^{3/2}e^{-m/T_F}\langle \sigma v \rangle M_{Pl}}{T_F^2} = 1 \quad . \tag{6.22}$$

Thus, the relic WIMP to entropy ratio is

$$\left(\frac{n_{eq}}{s}\right)_{T_F} \sim \frac{(mT_F)^{3/2}e^{-m/T_F}}{T_F^3} \sim \frac{1}{M_{Pl}m\langle \sigma v \rangle}\frac{m}{T_F} \quad . \tag{6.23}$$

For weakly interacting particles with masses in the GeV - TeV range, the ratio $m/T_F \sim 10 - 20$ at freeze-out. For concreteness, consider a massive neutrino (other WIMPs are qualitatively similar but differ in quantitative detail): for non-relativistic neutrinos of mass $m$, the four-momentum $p \sim m$, and the annihilation cross-section estimated at the beginning of the last section becomes $\sigma_\nu \sim G_F^2 m^2$. Substitution into (6.23) gives $n/s \sim 10^{-8}(m/\text{GeV})^{-3}$. Using eqns.(6.10), (6.12), and (6.16), this corresponds to a neutrino mass density of order $\Omega_\nu h^2 \simeq (m/\text{GeV})^{-2}$. This tells us that a particle with mass in the GeV range and annihilation cross-section typical of the weak interactions will have a relic density comparable to the closure density of the universe. Note that such cold particles have no trouble fitting into galaxy halos, unlike light neutrinos.

## 7. The Inflationary Universe

### 7.1. Motivation: the Horizon and Flatness problems

The inflationary scenario originally arose out of the attempt to solve several puzzles which arise when one extrapolates the standard cosmology back to the very early universe. The most important of these puzzles are the horizon and flatness problems, which we consider in turn.

Consider a light signal emitted from the origin at time $t_e$; from eqn.(2.10), it will reach a coordinate distance $r_h$ at a time $t$ given by

$$\int_0^{r_h} \frac{dr}{\sqrt{1-kr^2}} = \int_{t_e}^{t} \frac{dt}{a(t)} \quad . \tag{7.1}$$

If the second integral converges as $t_e \to 0$, then there exists a *particle horizon*: two fundamental observers separated by a coordinate distance larger than $r_h(t_e = 0)$ will not yet be in causal contact at time $t$. The proper distance from the origin to the coordinate $r_h(t_e = 0)$ at time $t$ is called the particle horizon radius $d_h$,

$$d_h(t) = a(t)\int_0^t \frac{dt}{a(t)} \quad . \tag{7.2}$$

In the standard cosmology, in the early radiation-dominated phase, $a(t) \propto t^{1/2}$, and the particle horizon $d_h \sim ct$. Note that this is also comparable to the Hubble length, $H^{-1}(t) \sim t$. More generally, $d_h$ is finite if the scale factor $a(t)$ grows more slowly than $t$ at early times; from (2.21), the criterion for this is $p > -\rho/3$, a reasonable condition for most fluids. Now recall that, in the absence of early reionization, the CMBR photons last scattered at a time $t_{dec} \simeq 5.6 \times 10^{12} h^{-1}$ sec. If $\Omega_0 \simeq 1$, photons emitted from two points separated by the particle horizon distance at the time of last scattering would arrive with an angular separation of order 1 degree at the observer today. Therefore, in the standard cosmology, when we look at a map of the microwave sky, we would expect it to look very anisotropic on angular scales larger than a degree, because we are seeing photons emitted from regions which were not yet in causal contact with each other at the time of last scattering. Yet the microwave sky is remarkably isotropic over all scales larger than a few arcminutes, to of order one part in $10^5$. This is the horizon problem: the CMBR is isotropic over scales which were not yet in causal contact when the photons last scattered. This argument might appear circular, for we have done this calculation using the homogeneous and isotropic FRW model, so by assumption the CMBR must appear isotropic over all scales, and one might conclude that there is no problem. This objection is spurious: we could easily repeat the derivation of the particle horizon in a cosmological model which is nearly but not exactly homogeneous and isotropic (say, with temperature fluctuations which are of order $10^{-2}$ instead of $10^{-5}$) and end up with essentially the same answer. That is, we do not need to assume exact homogeneity and isotropy for there to be a horizon problem.

The second puzzle of the standard cosmology is the flatness problem. In essence, this is the puzzle of why the spatial curvature term in (3.2) does not presently dominate over the matter-density term by a large margin: if $k \neq 0$ and $\lambda_0 = 0$, we can rewrite (3.2) as

$$\frac{1}{|\Omega(t) - 1|} = a^2(t) H^2(t) \quad . \tag{7.3}$$

From (2.21), $a^2 H^2 = \dot{a}^2$ is a decreasing function of time if $w \geq -1/3$. Consequently, in the standard cosmology (with non-negative fluid pressure), the observational fact that $\Omega_0$ is within an order of magnitude of unity means that $\Omega(t)$ must have been extremely close to unity at earlier times: at the Planck time, for example, it requires $|\Omega(t_{Pl}) - 1| \lesssim 10^{-60}$. In the absence of other prejudices, one might have guessed that the universe would emerge from the quantum gravity Planck era with the matter-density and curvature terms having comparable amplitudes, but the result above implies that they must have been identical at that time to one part in $10^{60}$. Another way to state this is: if conditions had been just slightly different at the Planck time, the universe would have gone into free expansion ($a \sim t$) or have recollapsed long before the present epoch. In other words, the natural timescale for the universe to become curvature-dominated is the Planck time, $t_{Pl} \sim 10^{-43}$ sec, yet our universe is still not strongly curvature-dominated at $t_0 \sim 10^{17}$ sec $\sim 10^{60} t_{Pl}$.

It is important to emphasize that these two puzzles are *not* inconsistencies of the standard cosmological model. Rather they point to features of the observed universe which the standard model does not explain and which moreover appear highly unlikely when the standard cosmology is embedded in a somewhat larger class of cosmological models. That is, they are both problems of initial conditions: in the "phase space" of initial conditions, the set of initial data that evolve to a Universe like ours is very small. A more physical way to say this is that homogeneity and spatial flatness are unstable properties of the model, so the present nearly homogeneous and spatially flat state of the Universe appears to be very sensitive to the initial conditions. Inflation was designed to reduce this sensitivity by widening the class of initial conditions which evolve to a nearly homogeneous and spatially flat state within the observable universe. This is the sense in which inflation is said to 'solve' the horizon and flatness problems.

### 7.2. Inflation: necessary ingredients

The discussion of the horizon and flatness problems implicitly pointed to their solution: what is wanted is an early era when the scale factor accelerates, $\ddot{a} > 0$, for this implies that $\dot{a}$ is increasing with time, i.e., that $a$ grows faster than $t$. Because of the rapid growth of the scale factor when it is accelerating, this was dubbed 'inflation' by Guth[49]. From (2.13), acceleration requires that the energy density be dominated by some fluid with equation of state $p < -\rho/3$. How long must such an inflationary epoch last? Suppose the accelerated expansion begins at an epoch $t_i$ and ends at $t_e$. Then, from (7.3), we can compare the present and initial curvature terms,

$$\frac{|\Omega_0 - 1|}{|\Omega_i - 1|} = \left(\frac{a_i H_i}{a_e H_e}\right)^2 \left(\frac{a_e H_e}{a_0 H_0}\right)^2 \simeq e^{2(60-N_e)} \quad . \tag{7.4}$$

In the last expression, I have assumed for the sake of argument that the universe is radiation-dominated for $t > t_e$, that the inflation epoch ends around the GUT scale, $T_e \sim 10^{15}$ GeV, and that $H_e \sim H_i$. Here, $N_e$ is the number of e-folds of growth of the scale factor during inflation,

$$a(t_e)/a(t_i) = \exp \int_{t_i}^{t_e} H(t) dt = e^{N_e} \quad . \tag{7.5}$$

We would like the ratio on the left hand side of (7.4) to be of order unity, so that $\Omega_i$ does not have to be extremely close to one. The right hand side of (7.4) then implies that the scale factor must grow by at least $N_e \simeq 60$ e-folds during the inflationary epoch. If $\Omega_i \sim 1$, this is identical to the condition that our observable universe have emerged from a region that was causally connected at the onset of inflation–thus solving the horizon problem as well.

The pleasing feature of inflation is that it drives the spatial curvature to zero via rapid expansion of the spatial hypersurfaces. After inflation, the curvature term again grows in importance, but it has been reset by inflation to such a tiny value that it is still catching up to the matter term today. In fact, if the universe enters an extended inflationary period, it is quite plausible that it lasts considerably longer than the required 60 e-folds of expansion, in which case the spatial curvature term would still be exponentially small today. In other words, unless $N_e$ is almost exactly 60, a sufficiently long period of inflation implies that the present universe should be observationally indistinguishable from being spatially flat, i.e., $\Omega_0 + \lambda_0 = 1$.

There is a second necessary ingredient for inflation to be a viable scenario for the very early universe. The accelerated expansion not only drives the spatial curvature to zero, but it also drives the density of matter and radiation to zero. At the end of $N_e \geq 60$ e-folds of expansion, the radiation temperature would be at most $T_e = T_i e^{-60}$ which is less than 100 eV if $T_i \leq M_{Pl}$. This would invalidate the argument in eqn.(7.4), where we assumed $T_e \sim M_{GUT}$, and, among other things, make it impossible to account for the light element abundances through primordial nucleosynthesis. Thus, a period of accelerated expansion by itself is not enough to solve the horizon and flatness problems: at the end of inflation, the energy density in the fluid that drives the inflationary expansion must be converted into radiation, to repopulate and reheat the universe. A minimal condition for successful inflation is that the reheat temperature after inflation be high enough for the baryon asymmetry to be generated; if baryogenesis takes place at the electroweak scale, this implies $T_e > 100$ GeV.

### 7.3. Inflation: Scalar Field Dynamics

In 1980, Guth[49] proposed that a scalar field trapped in a 'false vacuum' state with non-zero potential energy could act as the driving force of inflation and thereby solve the horizon and flatness problems. It is easy to see that this fits the bill for accelerated expansion: the energy-momentum components for a classical scalar field $\phi(\mathbf{x}, t)$ with potential $V(\phi)$ are

$$p_\phi = \frac{\dot\phi^2}{2} - \frac{(\nabla\phi)^2}{6} - V(\phi) \qquad (7.6)$$

and

$$\rho_\phi = \frac{\dot\phi^2}{2} + \frac{(\nabla\phi)^2}{6} + V(\phi) \quad . \qquad (7.7)$$

If the field is trapped in a local minimum $\phi_c$ of the potential, it will relax to a static configuration with $p_\phi = -\rho_\phi = -V(\phi_c) = $ constant, which satisfies the criterion for accelerated expansion. In this case, eqn.(2.12) becomes $H^2 = 8\pi G V(\phi_c)/3$ = constant, and the solution is the exponential de Sitter expansion of (3.7). In other words, a field dominated by constant potential energy acts as an effective cosmological constant.

While a promising idea, 'false vacuum' inflation did not satisfactorily incorporate the second necessary ingredient of reheating: once the field is trapped for sufficiently long in the false vacuum state, it must tunnel through the potential barrier which held it there, in order to reach the true vacuum where $V(\phi_t) = 0$. In this case, all the potential energy of the false vacuum ends up in the walls of the nucleated bubbles of true vacuum, and is not converted into radiation by bubble collisions efficiently enough. The remedy was soon provided by Linde[50] and by Albrecht and Steinhardt[51]: instead of the field being trapped in a metastable minimum of the potential, it can be classically evolving on a very flat potential if it is initially displaced from the potential minimum. If the potential is sufficiently flat, the kinetic and potential energy terms in eqns.(7.6-7) will redshift away with the expansion, leaving the potential term to dominate as before. Inflation ends when the field reaches a steeper part of the potential, the field speeds up and eventually oscillates about its potential minimum. If $\phi$ is coupled to lighter particles, these coherent field oscillations lead to particle creation and thereby reheat the universe: the scalar field energy is converted to radiation. This 'new inflation' model thus solves the 'graceful exit' problem of the original false vacuum scenario.

In these early inflation models, it was assumed that the scalar field driving inflation, the *inflaton*, was a Higgs field associated with the spontaneous breakdown of the grand unified gauge symmetry. The field would be initially displaced from its global minimum by finite temperature effects. It was soon realized, however, that the inflaton must be extremely weakly self-coupled in order for its quantum fluctuations to generate an acceptable amplitude of density perturbations. Since Higgs fields are coupled to gauge fields, however, radiative corrections typically generate large self-couplings. As a result, the concept of inflation was divorced from Higgs fields and gauge symmetry breaking, and a pandora's box of particle physics models was opened up. Nevertheless, the different models of inflation share many common dynamical features, and it is those which I will focus on.

To see how inflation works in a little more detail, consider the dynamics of the evolving scalar field. Within each Hubble volume, (*i.e.*, ignoring spatial gradients) the evolution of the field is described by the classical equation of motion for a homogeneous field $\phi(t)$,

$$\ddot{\phi} + 3H\dot{\phi} + \Gamma\dot{\phi} + V'(\phi) = 0 , \qquad (7.8)$$

where $\Gamma$ is the decay width of the inflaton field, a phenomenological term which has been introduced to roughly model the back-reaction effects of particle creation and reheating on the scalar field. The expansion rate $H = \dot{a}/a$ is determined by the Friedmann equation,

$$H^2 = \frac{8\pi}{3M_{Pl}^2}\left[V(\phi) + \frac{1}{2}\dot{\phi}^2\right] , \qquad (7.9)$$

and it is also useful to have the second order Friedmann equation,

$$\frac{\ddot{a}}{a} = -\frac{8\pi}{3M_{Pl}^2}\left[\dot{\phi}^2 - V(\phi)\right] . \qquad (7.10)$$

From (7.10), the condition that the universe be inflating, $\ddot{a} > 0$, requires that the kinetic energy by sub-dominant, $\dot{\phi}^2 < V$. A sufficient, although not necessary, condition for this is that the field be *slowly rolling* (SR) in its potential. In this context, 'slowly rolling' is a technical term: the field is said to be slowly rolling when its motion is overdamped, *i.e.*,

$$\ddot{\phi} \ll 3H\dot{\phi} \ , \qquad (7.11)$$

so that the $\ddot{\phi}$ term can be dropped in the equation of motion (7.8) (we assume $\Gamma \ll H$ during this phase). From (7.8), the defining SR condition implies that $\dot{\phi}^2 \ll 2V(\phi)$; thus, if the SR condition is well satisfied, the universe is inflating.

Slow-rollover is analytically convenient, for it implies two consistency conditions on the slope and curvature of the potential:

$$|V''(\phi)| \lesssim 9H^2 \text{ and } \left|\frac{V'(\phi)M_{Pl}}{V(\phi)}\right| \lesssim \sqrt{48\pi} \ , \qquad (7.12)$$

where the slow-roll equation of motion has been used in the last equality (eqn.(7.8) without the second derivativer term). Suppose inflation begins when the field value is $\phi(t_i) = \phi_i$, and the SR epoch ends when $\phi$ reaches a value $\phi_e$, at which one of the inequalities (7.12) is violated. To solve the cosmological puzzles, we demand that the scale factor of the universe inflates by at least 60 e-foldings during the SR regime,

$$N(\phi_i, \phi_e) \equiv \ln(a_e/a_i) = \int_{t_i}^{t_e} H dt = \frac{-8\pi}{M_{Pl}^2} \int_{\phi_i}^{\phi_e} \frac{V(\phi)}{V'(\phi)} d\phi \geq 60 \ . \qquad (7.13)$$

Once the potential is chosen, (7.12) determines $\phi_e$ in terms of the parameters in $V$. The condition (7.13) then constrains the initial value $\phi_i$ of the field.

Once $\phi$ grows beyond $\phi_e$, the SR condition breaks down, and the field evolution is more appropriately described in terms of oscillations about the potential minimum. These coherent oscillations excite the creation of particles to which $\phi$ is coupled, which in turn damps the oscillations and reheats the universe once the created particles thermalize by scattering. By energy conservation, the phenomenological damping term $\Gamma$ in the scalar equation of motion–the scalar decay rate–acts as a source for radiation,

$$\dot{\rho}_r + 4H\rho_r = \Gamma \dot{\phi}^2 \ . \qquad (7.14)$$

We can identify two qualitative regimes for reheating: (i) if $\Gamma > H$, reheating is very efficient: nearly all the scalar potential energy is converted to particles, leading to a reheat temperature given by $T_{RH}^4 \simeq V(\phi_e)$; (ii) on the other hand, if $\Gamma < H$, reheating is inefficient, and the scalar field oscillations partially redshift away before converting to particles. In this case, the reheating temperature is lower, and is given by

$$T_{RH} = (45/4\pi^3 g_*)^{1/4} \sqrt{\Gamma M_{Pl}} \ , \qquad (7.15)$$

where $g_*$ is the number of relativistic degrees of freedom. Clearly the reheating efficiency is determined by the couplings of the inflaton to other fields. Efficient reheating, say, with a reheat temperature high enough to allow baryogenesis at the GUT scale, $T_{RH} \sim 10^{14}$ GeV, requires relatively strong couplings. In many models, however, such couplings induce loop corrections to the scalar self-interaction which would upset the requisite flatness of its potential. As a result, a high reheat temperature is not always possible.

### 7.4. Quantum Fluctuations and Density Perturbations

Soon after the arrival of false vacuum inflation, it was appreciated that inflation could in principle[52,53] provide one of the holy grails of cosmology: a causal mechanism for the origin of density fluctuations that later grow to form large-scale structure. The problem can be seen by first considering the standard cosmology without inflation: in this case, the physical wavelength associated with a density perturbation mode grows with the scale factor, $\lambda_{phys} \sim a(t)$, while the Hubble length scales up more rapidly, $H^{-1} \sim t$ (recall the current Hubble distance is $d_h \sim H_0^{-1} = 3000h^{-1}$ Mpc). Consequently, at a redshift larger than $z_\lambda$, where $1 + z_\lambda \sim (H_0^{-1}/\lambda_0)^2$, perturbations with present scale $\lambda_0$ were larger than the instantaneous Hubble radius, which sets the lengthscale for causal processes. For perturbations on cosmologically interesting scales today, say, $\lambda_0 \sim 1 - 3000h^{-1}$ Mpc, this implies $z_\lambda \lesssim 10^6$, or $T_\lambda \lesssim 100$ eV (here, for the sake of simplicity, I have assumed matter-dominated expansion throughout; the reader can insert the appropriate radiation-dominated corrections to make the numbers more rigorous). In the standard cosmology, however, curvature perturbations cannot be causally generated on scales larger than the instantaneous Hubble radius, because of local energy conservation. Thus, perturbations on scales of galaxies and larger must have been generated by some physical process acting at $T \lesssim 100$ eV, after these scales crossed inside the Hubble radius. While some work has been done on a 'late time phase transition' at such a scale,[54] such models do require new physics at surprisingly low energy scales, and have not been met with overwhelming enthusiasm. Moreover, the resulting CMBR anisotropies in such scenarios may be intolerably large[55]. If one does not postulate a late-time origin for large-scale perturbations, then they must simply be postulated as initial conditions, because they could not have had a causal origin at epochs $z \gtrsim z_\lambda$. (Note that this argument applies to curvature perturbations; an important loophole is the early generation of large-scale isocurvature perturbations.)

An early epoch of inflation alters this situation: during accelerated expansion, the Hubble radius $H^{-1}$ grows more *slowly* than the scale factor (in a de Sitter epoch, $H^{-1}$ is constant). Thus, a perturbation of comoving wavelength $\lambda$ could be created causally on a physical scale less than $H^{-1}$ during inflation, expand outside the Hubble radius at some time $t_A(\lambda)$ during inflation, and then eventually 're-enter' the Hubble radius in the recent radiation- or matter-dominated era at a time $t_B(\lambda)$.

This hypothetical possibility was turned into a physical plausibility when it was realized that quantum fluctuations of the slowly rolling field in new inflation could provide a mechanism for generating density perturbations[56]. In a nutshell, one treats the spatial fluctuations of the field about its homogeneous classical expec-tation value as an approximately massless quantum field in de Sitter space; in its vacuum state, such a field has a spectrum of zero-point fluctuations, with an rms amplitude $\Delta\phi = H/2\pi$. These spatial variations in $\phi$ correspond to fluctuations in the inflaton energy density, which are converted into adiabatic density fluctuations in all species during reheating. The resulting density perturbation amplitude when a comoving wavelength $\lambda$ re-enters the Hubble radius is given approximately by

$$\left(\frac{\delta\rho}{\rho}\right)_{t_B(\lambda)} \simeq \left(\frac{H^2}{\dot{\phi}}\right)_{t_A(\lambda)} . \qquad (7.16)$$

It is straightforward to show that scales corresponding to the range from the galaxy scale to the present Hubble radius, $\lambda_0 = 1-3000h^{-1}$ Mpc, crossed outside the Hubble radius in the interval 50 - 60 e-folds before the end of inflation. During slow-rollover, the inflaton is effectively at its terminal velocity, which changes slowly over time; since the period of ten e-folds of the scale factor is brief, the right hand side of (7.16) changes very little over this interval. As a result, the density perturbation amplitude at Hubble radius crossing $(t_B(\lambda))$ is nearly constant.

Now, the density perturbation amplitude at Hubble radius crossing is essentially the gravitational potential perturbation on that scale, $\delta\Phi_\lambda \sim G\delta M_\lambda/\lambda \sim G\delta\rho_\lambda\lambda^2 \sim (\delta\rho/\rho)_\lambda(\lambda_{phys}/H^{-1})^2$, where we have used the Friedmann equation to relate $G\rho \sim H^2$. Thus, inflation predicts that the gravitational potential fluctuations are nearly independent of scale, $\delta\Phi_\lambda \simeq$ constant. Such a scale-invariant spectrum of potential fluctuations was first proposed on different grounds by Harrison, Zel'dovich, and Peebles and Yu, ten years before inflation. Note that the gravitational potential also sets the scale for the large-angle CMBR anisotropy through the Sachs-Wolfe effect: photons climbing out of deeper potential wells at the time of last scattering suffer a larger gravitational redshift, resulting in the temperature shift $\delta T/T = \delta\Phi/3 \sim (\delta\rho/\rho)_{\lambda=H^{-1}}$ (this result holds for the Einstein-de Sitter model, $\Omega_0 = 1$). Considerable excitement was generated when analysis of maps of the CMBR sky from the first year of COBE DMR data yielded a large-angle anisotropy consistent with the scale-invariant spectrum[3].

To get a feel for the result (7.16), I sketch out the steps of the calculation here (for a particularly complete discussion, see[57]). The idea is to treat small fluctuations semi-classically, by expanding the scalar field and metric perturbatively around their homogeneous solutions, e.g.,

$$\phi(\mathbf{x}, t) = \phi_o(t) + \delta\phi(\mathbf{x}, t) , \qquad (7.17)$$

where $\phi_o(t)$ is the classical homogeneous solution to the scalar equation of motion in the FRW background, eqn.(7.8), and $\delta\phi$ is a quantum field operator with equal-time commutation relation

$$[\delta\phi(\mathbf{x}, t), \partial\delta\phi(\mathbf{x}', t)/\partial t] = i\delta^3(\mathbf{x} - \mathbf{x}')/a^3(t) . \qquad (7.18)$$

A similar expansion is performed for the metric. It is convenient to consider the

Fourier transform

$$\delta\phi(\mathbf{x},t) = \int \frac{d^3k}{(2\pi)^3} \left[ a_{\mathbf{k}} \psi_k(t) e^{i\mathbf{k}\cdot\mathbf{x}} + \text{h.c.} \right] \quad , \tag{7.19}$$

where $a_{\mathbf{k}}$ is the annihilation operator for mode $\mathbf{k}$, $\psi_k(t)$ is the classical mode function, and $k = a k_{phys}$ is the (fixed) comoving wavenumber of the mode. We will consider the power spectrum $\mathcal{P}_\phi(k)$ of the field, the contribution per logarithmic wavenumber interval to the variance of the fluctuation,

$$\langle (\delta\phi)^2 \rangle = \int \frac{dk}{k} \mathcal{P}_\phi(k) \quad . \tag{7.20}$$

The power spectrum is proportional to the Fourier transform of the two-point correlation function of the field in the vacuum,

$$\mathcal{P}_\phi(k) = \frac{k^3}{2\pi^2} \int d^3x e^{i\mathbf{k}\cdot\mathbf{x}} \langle 0|\delta\phi(\mathbf{x})\delta\phi(0)|0\rangle = \frac{k^3}{2\pi^2} |\psi_k|^2 \quad . \tag{7.21}$$

In general the mode equation for $\psi_k$ can be complicated, since it involves coupled scalar and gravitational degrees of freedom. However, at least in the synchronous gauge ($\delta g_{00} = 0$), the term involving metric perturbations can be neglected, yielding

$$\ddot{\psi}_k + 3H\dot{\psi}_k + \frac{k^2}{a^2}\psi_k + V''(\phi_o)\psi_k = 0 \quad . \tag{7.22}$$

The slow rollover conditions imply that the potential term in (7.22) is also negligible, reducing the equation of motion to that of a free massless field in de Sitter space. The normalized solution is

$$\psi_k(t) = \begin{cases} \frac{\exp(ik \int dt/a)}{\sqrt{2ka}} & \text{for } k/aH \gg 1 \\ \text{constant} & \text{for } k/aH \ll 1 \end{cases} \quad . \tag{7.23}$$

The mode crosses outside the Hubble radius during inflation when $\lambda_{phys} = H^{-1}$, i.e., when $k = aH$. The solution (7.23) states that the mode oscillates inside the Hubble radius and then freezes out when it crosses outside. This is the mathematical statement of the physical intuition that microphysics only operates coherently on scales less than Hubble radius. For modes inside the Hubble radius, from (7.21) and (7.23) the power spectrum can be written $\mathcal{P}_\phi(k) = (H/2\pi)^2 (k/Ha)^2$. When a mode crosses outside the Hubble radius, the rms fluctuation on that scale is given by

$$(\Delta\phi)_{k=aH} = \mathcal{P}_\phi^{1/2}(k = aH) = \frac{H}{2\pi} \quad . \tag{7.24}$$

To find the resulting density perturbation amplitude, one must solve the perturbed Einstein equations; due to the gauge freedom in the choice of time coordinate, that is, in the way spacetime can be sliced up into spacelike hypersurfaces, one chooses a gauge to simplify the problem at hand. It is particularly convenient to make use of a gauge-invariant measure of the perturbation amplitude for adiabatic (curvature) perturbations,

$$\zeta = 3\Phi + \frac{\delta\rho}{p_o + \rho_o} \quad , \tag{7.25}$$

where $\Phi$ is an analog of the Newtonian potential that completely specifies the intrinsic Ricci curvature of the perturbed 3-surfaces. The variable $\zeta$ is useful because it generally satisfies $\zeta =$ constant for perturbations far outside the Hubble radius, $k \ll aH$, both during and after inflation. Moreover, the potential $\Phi$ is negligible at Hubble radius crossing, so, defining the power spectrum of $\zeta$ by analogy with (7.20), we have

$$\left(\frac{\delta\rho}{\rho}\right)_{t_B(\lambda)} \sim \mathcal{P}_\zeta^{1/2}(k) = \frac{(\Delta\phi)_{k=aH}V'(\phi_o)}{\dot\phi^2} = \left(\frac{3H^2}{2\pi\dot\phi}\right)_{t_A(\lambda)} \quad , \tag{7.26}$$

where we have used (7.24) and the slow-roll equation of motion, $3H\dot\phi = -V'(\phi)$. This completes the sketch of the derivation of (7.16).

Using the slow-roll equation of motion, the perturbation amplitude can be expressed in terms of the height and slope of the potential,

$$\mathcal{P}_\zeta^{1/2} = \sqrt{384\pi}\frac{V^{3/2}}{V'M_{Pl}^3} \quad . \tag{7.27}$$

A useful way to characterize the scale-dependence of the perturbations is to consider the density perturbation power spectrum, $P_\rho(k) = |\delta_k|^2$, where $\delta_k$ is the Fourier transform of $\delta\rho(\mathbf{x})/\bar\rho$, and we relate different conventions for what is called the 'power spectrum' by using the notation $\mathcal{P} = k^3 P$. The density spectrum can be written as a power law in wavenumber, $P_\rho(k) \sim k^{n_s}$, where $n_s = 1$ corresponds to the scale-invariant spectrum, i.e., to $\mathcal{P}_\zeta =$ constant. (To see this, recall that $\delta\Phi_\lambda \sim (\delta\rho/\rho)_\lambda(\lambda/H^{-1})^2$, and that the contribution per logarithmic wavenumber interval to the rms fluctuation on scale $\lambda$ is $(\delta\rho/\rho)_\lambda \sim k^{3/2}|\delta_k|$.) Small deviations from scale-invariance can be expressed in terms of the index $n_s - 1$.

One can carry through an argument similar to that above for gravitational wave perturbations in de Sitter space. Expanding the metric about the de Sitter FRW solution, the tensor modes $h_{+,\times}$ satisfy the massless scalar field equation (7.22) (with no potential term). Defining a canonically normalized scalar field for each polarization state, $h_{+,\times} = \sqrt{16\pi G}\phi_{+,\times}$, the tensor amplitude at Hubble crossing is

$$h_{k=Ha} \simeq \frac{2}{\sqrt\pi}\frac{H}{M_{Pl}} \simeq \frac{V^{1/2}}{M_{Pl}^2} \quad . \tag{7.28}$$

The induced Sachs-Wolfe anisotropy on large scales is $\delta T/T \sim h_{k=Ha}$.

## 7.5. Chaotic Inflation: a worked example

The result (7.26) for the perturbation amplitude provides a severe constraint on the form of the inflaton potential. To get a feeling for the numbers, consider one of the simplest forms for the potential, a pure quartic $V(\phi) = \lambda \phi^4/4!$, an example of Linde's chaotic inflation. In this case, the field rolls down to the origin from its initial value; the slow roll conditions (7.12) are satisfied for $\phi > \phi_e = M_{Pl}/\sqrt{2\pi}$, and the number of inflation e-folds is given by (7.13),

$$N_e(\phi_i, \phi_e) = \frac{2\pi}{M_{Pl}^2}\left(\phi_i^2 - \phi_e^2\right) = \frac{2\pi \phi_i^2}{M_{Pl}^2} - 1 \ . \tag{7.29}$$

Thus, the condition of sufficient inflation, $N_e > 60$, requires $\phi_i > 3.1 M_{Pl}$, and the density perturbation amplitude on the scale of the present Hubble radius is

$$P_\zeta^{1/2} = (\pi \lambda)^{1/2} \left(\frac{\phi_i}{M_{Pl}}\right)^3 \simeq 50 \lambda^{1/2} \ . \tag{7.30}$$

The perturbation spectrum is almost exactly scale-invariant in this model, $n_s \simeq 0.95$, because $\phi$ changes very little in the interval between 60 and 50 e-folds before the end of inflation. The COBE observation of CMBR anisotropy indicates $P_\zeta^{1/2} \simeq 10^{-4}$, which implies the inflaton self-coupling must satisfy

$$\lambda \leq 4 \times 10^{-12} \ . \tag{7.31}$$

## 7.6. Naturalness and the Small Coupling Problem

The preceding example illustrates a general conclusion that applies to all inflation models: the inflaton field must be very weakly self-coupled. In different models, this constraint appears in different guises, but it is alway present in one form or another: the potential must contain a very small dimensionless number of order $10^{-12}$.

Attitudes concerning this problem vary widely among inflation theorists: it is often said that $\lambda_\phi \sim 10^{-12}$ constitutes unacceptable 'fine tuning', but that is a misapplication of the fine tuning concept in this context. To others, it is not an issue of great concern, because we know there exist other small numbers in physics, such as lepton and quark Yukawa couplings $g_Y \sim 10^{-5}$ and the ratio $M_{weak}/M_{Pl} \sim 10^{-17}$. Partly as a consequence of the latter view, in recent years, it has become customary to decouple the inflaton completely from particle physics models, to specify an 'inflaton sector' with the requisite properties, with little regard for its physical origin.

Nevertheless, it is meaningful and important to ask whether such a small value for $\lambda_\phi$ is in principle unnatural. Clearly, the answer depends on the particle physics model within which $\phi$ is embedded and on one's interpretation of naturalness. A small parameter $\lambda$ is said to be "technically natural" if it is protected against large radiative corrections by a symmetry, *i.e.*, if setting $\lambda \to 0$ increases the symmetry of the system. For example, in this way, low energy supersymmetry might protect the small ratio $M_{weak}/M_{Pl}$ by cancelling boson and fermion loops. However, in technically natural models, the small coupling $\lambda_\phi$, while stable against radiative corrections, is itself unexplained, and is generally postulated (*i.e.*, put in by hand) solely in order to generate successful inflation. Technical naturalness is a useful concept for low energy effective Lagrangians, like the electroweak theory and its supersymmetric extensions, but it points to a more fundamental level of theory for its origin. For example, the underlying origin of the mass hierarchy $M_{weak}/M_{Pl}$ in supersymmetric theories is thought to be associated with hidden sector physics in supergravity theories, which start out with no small dimensionless couplings at the Planck scale. Since inflation takes place *relatively* close to the Planck scale, it would be preferable to find the inflaton in such particle physics models which are "strongly natural", that is, which have no small numbers in the fundamental Lagrangian.

In a strongly natural gauge theory, all small dimensionless parameters ultimately arise dynamically, *e.g.*, from renormalization group (or instanton) factors like $\exp(-1/\alpha)$, where $\alpha$ is a gauge coupling. In particular, in an asymptotically free theory, the scale $M_1$, at which a logarithmically running coupling constant becomes unity, is $M_1 \sim M_2 e^{-1/\alpha}$, where $M_2$ is the fundamental mass scale in the theory. In some models, the inflaton coupling $\lambda_\phi$ arises from a ratio of mass scales, $\lambda_\phi \sim (M_1/M_2)^q$. As a result, in such models, $\lambda_\phi$ is naturally exponentially suppressed, $\lambda_\phi \sim e^{-q/\alpha}$.

One example of a class of such models is 'natural inflation',[58,59] in which the role of the inflaton is played by a pseudo-Nambu-Goldstone-boson (in particle physics models, an example of this is the axion). In its simplest incarnation, consider a global $U(1)$ symmetry spontaneously broken at a high-energy scale $f$ by a non-zero expectation value for a complex scalar field, the potential for which takes the familiar Mexican-hat (or wine bottle) form. At scales much below $f$, the only relevant degree of freedom is the massless angular variable $\phi$ around the bottom of the potential. At an energy scale $\Lambda \ll f$, the symmetry is explicitly broken–the Mexican-hat is tilted, generating a potential for $\phi$ which is generally of the form

$$V(\phi) = \Lambda^4[1 \pm \cos(N\phi/f)] \ . \qquad (7.32)$$

In axion models, $\Lambda$ characterizes the scale at which a running gauge coupling becomes strong, and is related to the fundamental scale $f$ by $\Lambda \sim f e^{-q\alpha}$. For $f \sim M_{Pl} \sim 10^{19}$ GeV and $\Lambda \sim M_{GUT} \sim 10^{15}$ GeV, this field can drive inflation; in this case, the effective quartic coupling is $\lambda_\phi \sim (\Lambda/f)^4 \sim 10^{-13}$, as required. In the standard model and in typical grand unified theories, the only strong gauge coupling is that of QCD, with $\Lambda_{QCD} \simeq 100$ MeV. However, in supergravity and superstring models, there is a 'hidden' gauge sector which interacts only gravitationally with ordinary quarks and leptons. It has been suggested that the hidden sector gauge interactions become strong at a scale comparable to that above, $\Lambda \sim 10^{14}$ GeV, in order to break supersymmetry at the TeV scale in the observable sector.

An interesting possibility in this case is that the perturbation spectrum can deviate significantly from scale-invariance: the spectral index is given by

$$n_s \simeq 1 - \frac{M_{Pl}^2}{8\pi f^2} \quad . \tag{7.33}$$

The constraint that the universe reheat after inflation to a temperature $T_{RH} \gtrsim 100$ GeV leads to the constraint $f/M_{Pl} > 0.3$, which implies $n_s \gtrsim 0.5$, allowing a small but significant break from scale-invariance ($n_s = 1$). Contrary to some erroneous statements in the literature, such a breaking from scale-invariance does *not* require any fine-tuning of the parameters in this model. Arguments from the formation of structure, combined with the COBE observations, at any rate indicate that the spectral index $n_s \gtrsim 0.6 - 0.7$.

## 8. A Brief Look at Structure Formation

Let us finish with a cursory look at how primordial perturbations, e.g., from inflation, end up looking today. As discussed above, the *primordial* spectrum from inflation is generally a power-law in wavenumber, $P_\rho(k) = |\delta_k(t_i)|^2 = Ak^{n_s}$, and the present spectrum is related to the primordial one through a transfer function, $|\delta_k(t_0)|^2 = T^2(k)|\delta_k(t_i)|^2$. The transfer function $T(k)$ encodes the scale-dependence of the linear ($\delta \ll 1$) gravitational evolution of the perturbation modes; it depends on the nature of the dark matter (hot or cold), its density ($\Omega_{DM}$) as well as $\Omega_B$ and $H_0$. On scales which enter the Hubble radius after the universe becomes matter-dominated, $k \ll k_{eq} \simeq 0.2\Omega_0 h^2$ Mpc$^{-1}$, all perturbations undergo the same growth rate, so the transfer function $T^2(k) \simeq 1$; on smaller scales, $k \gg k_{eq}$, it bends over to $T^2(k) \sim k^{-4}$ as $k \to \infty$, reflecting the suppressed growth of fluctuations which cross inside the Hubble radius while the universe is still radiation-dominated. For standard cold dark matter (CDM), we have $h = 0.5$ and assume negligible baryon density, $\Omega_B \ll 1$, $\Omega_{cold} = 1$, leading to the characteristic scale $k_{eq} \simeq 0.05$ Mpc$^{-1}$. A reasonable analytic fit to the linear transfer function for CDM models is given by[60]

$$T(k) = \left(1 + \frac{1.7k}{\Omega h} + \frac{9k^{3/2}}{(\Omega h)^{3/2}} + \frac{k^2}{(\Omega h)^2}\right)^{-1} \quad , \tag{8.1}$$

(in the approximation $\Omega_B = 0$, $\Omega = \Omega_{cold}$). For hot dark matter, on the other hand, in the absence of seed perturbations such as cosmic strings, the transfer function is also exponentially damped by relativistic neutrino free-streaming for wavenumbers larger than $k_\nu \simeq 0.1(m_\nu/20\text{eV})$ Mpc$^{-1}$. Finally, the present density spectrum is related to the galaxy power spectrum by a bias prescription; for the simplest linear bias model,

$$P_{gal}(k) = b_{gal}^2 P_\rho(k) = b_{gal}^2 T^2(k)|\delta_k(t_i)|^2 \quad . \tag{8.2}$$

As mentioned in Sec. 2, galaxy catalogs such as the CfA, the IRAS 1.2 Jy and QDOT, and the angular APM and EDSGC catalogs, suggest that the spectrum predicted in (8.1) and (8.2) with standard CDM ($\Omega h = 0.5$), linear bias, and $n_s = 1$ may not have exactly the right shape to match the observed galaxy spectrum, having relatively too little (too much) power on large (small) scales. Solutions to this extra-power problem can be classified according to which element on the right hand side of Eqn.(8.2) one tinkers with. In all cases, the aim is to flatten the shape of the spectrum at intermediate wavenumbers $k \sim 0.05$ h Mpc$^{-1}$ by increasing the relative power on these scales compared to smaller wavelengths. For example, as (8.1) indicates, one can abandon 'standard' CDM and increase the characteristic wavelength where the transfer function $T(k)$ bends down, providing more relative large-scale power, by reducing $\Omega_0$ from unity; in the context of inflation, this requires either a special choice of the number of e-folds $N_e$ or the introduction of a cosmological constant to ensure $\Omega_0 + \lambda_0 = 1$. An example of this is shown in Fig. 7, which again shows the APM galaxy angular correlation function plotted against the prediction for $w(\theta)/(b_{gal}\sigma_8)^2$ (see below) in linear theory for standard and low-density CDM, both with $n_s = 1$. Recall that lowering $\Omega_0$ also increases the expansion age $H_0 t_0$.

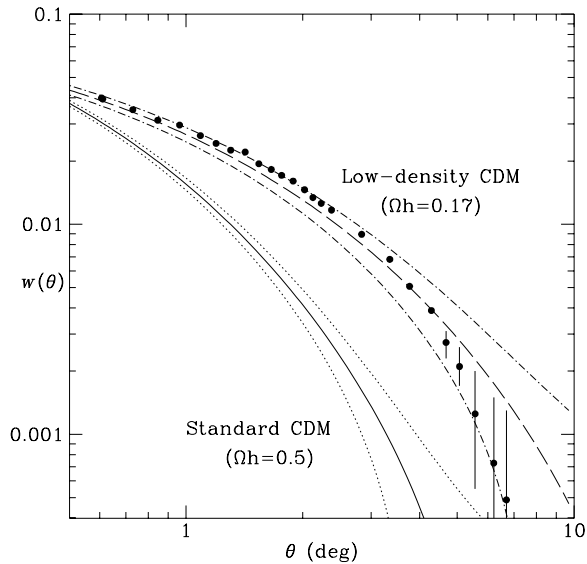

Figure 7: The APM galaxy angular correlation function of Fig. 1 is shown with two model predictions: standard ($\Omega h = 0.5$) CDM and a best-fit low-density CDM model ($\Omega h = 0.17$) (from ref. 62). The upper and lower bracketing curves for each model indicate the expected $1 - \sigma$ spread due to cosmic variance.

Another recently popular alternative is to employ a mixture of hot and cold dark matter, with $\Omega_{hot} \simeq 1 - \Omega_{cold} = 0.3$; from eqn.(6.21), this can be achieved with a light neutrino with mass $m_\nu \simeq 7$ eV. Due to free-streaming, the neutrino admixture partially suppresses power on small scales, which is what is wanted. A third possibility is to retain standard CDM but consider primordial perturbation spectra with $n_s < 1$ from inflation, but it appears that such tilting of the spectrum, when normalized to COBE, does not by itself solve the power problem[59] without violating other constraints. Alternatively, one can admit a more complex scheme for biasing in which the bias factor in (8.2) is scale-dependent, $b_g \to b_g(k)$, and increases at large scales[61,62]. Unlike the other fixes, in this case the extra power on large scales is an optical illusion, a property of the galaxy field but not the underlying density field. This difference leads to qualitatively different behavior of the higher-order (e.g., three-point) galaxy correlation functions, an effect which should be testable against observations[63].

Since all inflation models predict a deviation from scale-invariance at some level, it is interesting to see what limits on the spectral index $n_s$ follow from COBE in the context of standard CDM. For these models, the density power spectra form a two-parameter family characterized by the spectral index $n_s$ and the normalization $A$. Instead of $A$, it is common to normalize spectra by the *rms* linear mass fluctuation in spheres of radius 8 $h^{-1}$ Mpc, $\sigma_8 \equiv \langle(\delta M/M)^2\rangle^{1/2}_{R=8h^{-1}\rm Mpc}$, where

$$\sigma_R^2 = \frac{1}{2\pi^2}\int_0^\infty dk k^2 P(k) W^2(kR) \, , \tag{8.3}$$

and the window function

$$W(kR) = \frac{3}{(kR)^3}(\sin kR - kR \cos kR) \tag{8.4}$$

filters out the contribution from small scales. Redshift surveys of optically selected galaxies (in particular the CfA and APM surveys) indicate that the variance in galaxy counts on this scale is of order unity. Thus, in a linear bias model (and ignoring non-linear gravitational effects), the bias factor for these galaxies would be $b_{opt} \sim 1/\sigma_8$; for other galaxy populations, $b_{gal}\sigma_8$ may differ from unity.

From the Sachs-Wolfe effect, for a given normalization and spectral index, we can calculate the expected large-angle CMBR anisotropy. The rms temperature fluctuation on the scale of $10^o$ observed by COBE, $\sigma_T(10^o) = 1.085 \times 10^{-5}(1 \pm 0.169)$, then yields a relation between them, $\sigma_8 \simeq e^{-2.63(1-n_s)}$ [$1 \pm 0.2$]. Thus, models which obtain more relative large-scale power by tilting the spectrum to $n_s < 1$ reduce the power on galaxy scales. If $\sigma_8 \lesssim 0.5$, the power on small scales is reduced to the point that galaxy formation would occur too recently; in the context of these models, this leads to the constraint $n_s \gtrsim 0.6 - 0.7$. This bound assumes that gravitational waves make a negligible contribution to the CMBR signal. While this is a good approximation for models such as natural inflation, there are other inflation potentials for which the gravity wave contribution to the large-angle CMBR anisotropy becomes appreciable when $n_s < 1$. An example is power law inflation, produced by an exponential potential for $\phi$: in this case, the relative contribution of the tensor modes compared to the scalar (density perturbation) modes to the large-angle anisotropy is $R \simeq 6(1 - n_s)$. However, contrary to some statements in the literature, this relation is *not* generic to inflation, but only holds approximately for a restricted class of models. Clearly, for models in which it does apply, the effect is to lower the COBE constraint on the perturbation amplitude $\sigma_8$ for fixed $n_s$. For models with $n_s < 1$, this makes the epoch of structure formation more recent, and the corresponding lower bound on $n_s$ even tighter. As this subject has become a booming theoretical industry in the last year, I refer the reader to recent issues of Physical Review D for more details.

## Acknowledgements

I would like to thank the TASI organizers, T. Walker, S. Raby, and K. T. Mahanthappa for putting together an enjoyable school, the students for asking good questions without throwing anything, and the Chicago Bulls for experimentally verifying my prediction. This work was supported in part by the DOE and by NASA (grant NAGW-2381) at Fermilab.

## Appendix A: Notation and Units

It is useful to be able to translate back and forth between astronomical units and high energy physics units. The fundamental unit of distance in extragalactic astronomy is the parsec (pc): 1 pc = 3.26 light-years = $3.09 \times 10^{18}$ cm. The typical galactic scale is measured in kiloparsec (kpc): for example, the solar system is approximately $R_0 = 8 - 8.5$ kpc from the center of the Milky Way. The nearest large galaxy to our own is Andromeda, at a distance of roughly 700 kpc, and the typical distance between neighboring large galaxies (outside clusters) is a few Megaparsec (Mpc). Absolute extragalactic distances are uncertain by a factor of order 2, and are usually given in terms of $h^{-1}$ Mpc, where the Hubble parameter $H_0 = 100h$ km/sec/Mpc, and observations indicate $0.4 < h < 1$ (see below). Cosmological distances are often estimated using redshifts via the Hubble law, $v = cz = H_0 d$, and so are sometimes expressed in terms of recession velocity: 100 km/sec = 1 $h^{-1}$ Mpc. Astronomical masses and luminosities are measured in solar units, $M_\odot = 1.989 \times 10^{33}$ gm, $L_\odot = 3.86 \times 10^{33}$ erg/sec. Large galaxies commonly have a luminosity of order $10^{10} - 10^{11} L_\odot$ and mass of order $10^{11} - 10^{12} M_\odot$. In particle physics, it is often convenient to calculate in natural units, in which $\hbar = c = k_B = 1$. In this case, all dimensionful quantities can be expressed in units of energy, say in GeV; for example, Length = 1/Energy. At the end of a calculation, to convert an expression to physical units, insert appropriate powers of $\hbar c = 2 \times 10^{-14}$ GeV cm, and convert lengths to times by dividing by $c = 3 \times 10^{10}$ cm/sec. When gravity is involved, use the Planck mass (energy) $M_{Pl} = G^{-1/2} = 1.2 \times 10^{19}$ GeV. Finally, when converting energy to temperature ($E = k_B T$) and vice versa, use 1 eV = $1.16 \times 10^4$ K. To go from particle physics to astronomical units, it is sometimes handy to remember that the sun contains about $10^{57}$ protons, i.e., $10^{57}$ GeV $\simeq 1 M_\odot$.